\documentclass[printer]{aa}

\usepackage[utf8]{inputenc}
\usepackage{graphicx}
\usepackage{xcolor}
\usepackage[colorlinks=true,breaklinks=true,citecolor=blue,linkcolor=blue,urlcolor=blue]{hyperref}

\defcitealias{Franz-2020}{I}
\defcitealias{Franz-2022a}{II}

\begin{document}

\title{Dust entrainment in photoevaporative winds:\\Synthetic observations of transition disks}

\author{R.~Franz \inst{1}\thanks{\href{mailto:rfranz@usm.lmu.de}{rfranz@usm.lmu.de}} \and G.~Picogna \inst{1} \and B.~Ercolano \inst{1,2} \and S.~Casassus \inst{3} \and T.~Birnstiel \inst{1,2} \and Ch.~Rab \inst{1,4} \and S.~P{\'e}rez \inst{5,6}}

\institute{University Observatory, Faculty of Physics, Ludwig-Maximilians-Universit{\"a}t M{\"u}nchen, Scheinerstr.~1, 81679 Munich, Germany
\and
Excellence Cluster Origin and Structure of the Universe, Boltzmannstr.~2, 85748 Garching, Germany
\and
Departamento de Astronom{\'i}a, Universidad de Chile, Casilla 36-D, Santiago, Chile
\and
Max-Planck-Institut f{\"u}r extraterrestrische Physik, Giessenbachstr.~1, 85748 Garching, Germany
\and
Departamento de F{\'i}sica, Universidad de Santiago de Chile, Av. Ecuador 3493, Estaci{\'o}n Central, Santiago, Chile
\and
Center for Interdisciplinary Research in Astrophysics and Space Exploration (CIRAS), Universidad de Santiago de Chile, Chile
}

\date{Received 30 Nov 2021 / Accepted 25 Jan 2022}

\abstract
% Context:
{X-ray- and extreme-ultraviolet- (XEUV-) driven photoevaporative winds acting on protoplanetary disks around young T-Tauri stars may strongly impact disk evolution, affecting both gas and dust distributions.
Small dust grains in the disk are entrained in the outflow and may produce a detectable signal.
In this work, we investigate the possibility of detecting dusty outflows from transition disks with an inner cavity.}
% Aims:
{We compute dust densities for the wind regions of XEUV-irradiated transition disks and determine whether they can be observed at wavelengths $0.7 \lesssim \lambda_\mathrm{obs} \, [\mu\mathrm{m}] \lesssim 1.8$ with current instrumentation.}
% Methods:
{We simulated dust trajectories on top of 2D hydrodynamical gas models of two transition disks with inner holes of 20 and 30\,AU, irradiated by both X-ray and EUV spectra from a central T-Tauri star.
The trajectories and two different settling prescriptions for the dust distribution in the underlying disk were used to calculate wind density maps for individual grain sizes.
Finally, the resulting dust densities were converted to synthetic observations in scattered and polarised light.}
% Results:
{For an XEUV-driven outflow around a $M_* = 0.7\,\mathrm{M}_\odot$ T-Tauri star with $L_X = 2 \cdot 10^{30}\,\mathrm{erg/s}$, we find dust mass-loss rates $\dot{M}_\mathrm{dust} \lesssim 2.0 \cdot 10^{-3}\,\dot{M}_\mathrm{gas}$, and if we invoke vertical settling, the outflow is quite collimated.
The synthesised images exhibit a distinct chimney-like structure.
The relative intensity of the chimneys is low, but their detection may still be feasible with current instrumentation under optimal conditions.}
% Conclusions:
{Our results motivate observational campaigns aimed at the detection of dusty photoevaporative winds in transition disks using JWST NIRCam and SPHERE IRDIS.}


\maketitle

%%%%%%%%%%%%%%%%%%%%%%%%%%%%%%%%%%%%%%%%%%%%%%%%%%%%%%%%%%%%%%%%%%%%%%%%%%%%%%%%%%%%%%%%%%%%%%%%%%%%

\section{Introduction}
\label{sec:Intro}

Planets form from the reservoir of gas and dust within the protoplanetary disk initially surrounding the host star \citep[see e.g.][]{Armitage-2018}; once this material has been dispersed, planet formation necessarily comes to a halt.
Photoevaporative winds, driven by highly energetic radiation from the central star, are one of the mechanisms proposed to clear out disk material \citep[e.g.][]{Clarke-2001}; especially X-ray and extreme-ultraviolet (together: XEUV) winds are thought to be very efficient at driving outflows that eventually disperse the disk from the inside out, via the formation of a transition disk \citep{Ercolano-2009, Owen-2010, Owen-2012a}. Transition disks, that is disks with an inner cavity, may also be formed by other processes, such as dynamical interactions with giant planets or magneto-hydrodynamical (MHD) winds \citep[see e.g.][]{Kunitomo-2020, Pascucci-2020}.
The detection of photoevaporative winds is important in order to assess their role in disk evolution; so determining possible observational tracers is key to refine our understanding of this important mechanism.

The presence of disk winds can be inferred in a variety of ways; for instance, \citet{Monsch-2019, Monsch-2021a, Monsch-2021b} predicted orbital distributions of hot Jupiters in XEUV-irradiated disks, and \citet{Ercolano-2010}, \citet{Ercolano-2016}, and \citet{Weber-2020} modelled line profiles for disks impacted by photoevaporation.
Furthermore, \citet{Owen-2011a}, \citet{Hutchison-2016c, Hutchison-2016b}, \citet{Booth-2021a}, and \citet{Hutchison-2021} have worked to model and analytically formulate dust entrainment in photoevaporative winds.
In \citet[][henceforth Paper~I]{Franz-2020} and \citet[][henceforth Paper~II]{Franz-2022a}, we numerically simulated the dust content and observability of dusty XEUV-driven outflows around a primordial protoplanetary disk based on the hydrodynamical (HD) models of \citet{Picogna-2019} \citep[which have since been refined by][]{Ercolano-2021, Picogna-2021}.
The results of Papers~\citetalias{Franz-2020} and \citetalias{Franz-2022a} have shown that a dusty outflow signature is indeed expected at $\mu$m-wavelengths for the primordial disk model investigated there (hereafter `PD'), but its detection and interpretation would be challenging with current instrumentation.

In this work, we set out to investigate possible signatures of dust entrained in photoevaporative winds launched from transition disks, that is disks where an inner cavity has already formed.
In these objects, the stellar radiation reaches the disk midplane at the gap edge, allowing more material to be entrained from this location where gas and dust densities are much higher than at the disk-wind interface; the latter is several scale heights above the midplane \citep[see e.g.][]{Ercolano-2017b}.

% Structure of the paper
This paper is organised as follows:
The calculations to obtain the dust densities and synthetic observations are outlined in Sect.~\ref{sec:Methods}.
In Sect.~\ref{sec:Results}, the resulting density maps and observational concurrences are presented.
We discuss our findings in Sect.~\ref{sec:Discussion} and summarise them in Sect.~\ref{sec:Summary}.

%%%%%%%%%%%%%%%%%%%%%%%%%%%%%%%%%%%%%%%%%%%%%%%%%%%%%%%%%%%%%%%%%%%%%%%%%%%%%%%%%%%%%%%%%%%%%%%%%%%%

\section{Methods}
\label{sec:Methods}

Our goal is to investigate the observability of photoevaporative winds in transition disks as traced by the entrained dust grains.
To this end, we have followed the approach of Paper~\citetalias{Franz-2020} to first simulate dust grain trajectories, and of Paper~\citetalias{Franz-2022a} to then obtain dust density maps and synthetic observations.
The general methods employed will be briefly summarised in the next subsections, but we refer to Papers~\citetalias{Franz-2020} and \citetalias{Franz-2022a} for a more detailed description.

\subsection{Gas models}
\label{sec:Methods:gas}

We consider two model disks with cavities of $r_\mathrm{gap} \approx \lbrace 20, 30 \rbrace \,$AU, which we refer to as `TD20' and `TD30' below, respectively.
Significantly larger gap sizes lead to instabilities with the employed setup \citep[see][]{Woelfer-2019}.
Much smaller gap sizes would be difficult to observationally distinguish from primordial disks with current instruments; in addition, we wanted to investigate radii close to which the XEUV wind reaches its full potential for this model (see Paper~\citetalias{Franz-2020}).
This corresponds to the overall goal of Paper~\citetalias{Franz-2022a} to present a best-case scenario for the observability of a photoevaporative wind launched from the disk model.
As in Papers~\citetalias{Franz-2020} and \citetalias{Franz-2022a}, the gas models for the disks -- in this case, slightly modified versions of the transition disks presented by \citet{Picogna-2019} -- were computed using a modified version of the \texttt{Pluto} code \citep{Mignone-2007} for the hydrodynamical evolution.\footnote{\texttt{Pluto}: \href{http://plutocode.ph.unito.it/}{[link]}. Version 4.2 was used for this work.}
This modified version includes a temperature prescription obtained via \texttt{Mocassin} \citep{Ercolano-2003, Ercolano-2005, Ercolano-2008a}.
In these simulations, azimuthal and midplane symmetries were assumed as well as an $\alpha$-viscosity \citep{Shakura-1973} of $1 \cdot 10^{-3}$ \citep[to match the new setup of][]{Picogna-2021} and a mean atomic mass of $1.37125$.
The initial density distributions of \citet{Picogna-2019} were obtained from a steady-state primordial disk profile, adding an exponential cut-off function at the location of the gap.
The steady-state profile of the transition disks used in this work are the result of the interaction of the viscous spreading of the initial density profile, and the increased mass-loss rate due to the photoevaporative wind at the inner disk edge.

The system consists of a $M_* = 0.7\,\mathrm{M}_\odot$ T-Tauri star with an X-ray luminosity of $L_X = 2 \cdot 10^{30}\,\mathrm{erg/s}$ \citep[which represents the median of the stars in Orion, see][]{Preibisch-2005b}.
The mass of the primordial disks, prior to gap opening, was $M_\mathrm{disk} \simeq 0.01\,M_*$, of which about 60\% (TD20) or 45\% (TD30) remain in the transition disk models.
The corresponding photoevaporation-driven gas mass-loss rates are $\dot{M}_\mathrm{gas}^\mathrm{\,TD20} \simeq 3.0 \cdot 10^{-8} \,\mathrm{M_\odot/yr}$ and $\dot{M}_\mathrm{gas}^\mathrm{\,TD30} \simeq 2.9 \cdot 10^{-8} \,\mathrm{M_\odot/yr}$.\footnote{\citet{Picogna-2019} found $\dot{M}_\mathrm{gas}$ to be higher for transitional than for primordial disks. Our values of $\dot{M}_\mathrm{gas}$ are lower than for the PD of Paper~\citetalias{Franz-2022a} solely due to differences in the model setup. As laid out in Sect.~\ref{sec:Results:dust:Mdot}, this does not detrimentally affect our results.}

The region between the star and $r_\mathrm{gap}$ has been completely cleared in our models.
We do not consider transition disks with an inner disk \citep[see e.g.][]{Francis-2020}; in the case of an inner disk blocking the stellar radiation, one can assume that these objects behave similar to primordial disks, unless a strong difference in gas scale heights between the inner and outer disks exists.

\subsection{Dust in the wind}
\label{sec:Methods:dust}

To summarise the methodology of Paper~\citetalias{Franz-2022a}, the gas distribution in the disk \citep[taken from][see above]{Picogna-2019} was used to compute the dust distribution up to the disk-wind interface (the `base' of the wind or disk surface, determined as the location of the largest drop in gas temperature, see Paper~\citetalias{Franz-2020}).
There, the gas and dust densities ($\varrho_\mathrm{gas}$ and $\varrho_\mathrm{dust}$) are thus directly linked.
Then, a collection of dust grain trajectories was simulated using the prescriptions of \citet{Picogna-2018}; they were launched directly from the base in order to focus on the dust motion in the wind.
We did not use a physical density prescription for their initial placement.
This allowed us to derive normalised wind density maps from the trajectories via a particle-in-cell approach; these were combined with the base densities to obtain $\varrho_\mathrm{dust}$ in the wind.

\subsubsection{Trajectories}
\label{sec:Methods:dust:traj}

We employed a grid extending to $r \lesssim 300\,$AU, with $r = \sqrt{x^2+y^2+z^2}$ to model trajectories for eight distinct grain sizes $a_0$, in this case $a_0 \in \lbrace 0.01, 0.1, 0.5, 1, 2, 4, 8, 12 \rbrace \,\mu$m, with an internal grain density of $\varrho_\mathrm{grain} = 1.0 \,\mathrm{g/cm}^3$ (as in Papers~\citetalias{Franz-2020} and \citetalias{Franz-2022a}).
The grains were initially placed along the disk surface within $r_\mathrm{gap} \lesssim r \lesssim 200\,$AU.
To fully cover the strong vertical slope close to the outer gap edge,\footnote{With `outer gap edge', we refer to the outer edge of the (inner) cavity.} we placed the particles along the $(r,\vartheta)$-coordinates of the disk surface using a uniform random distribution.

The simulations were evolved for about $3.8\,$kyr; a constant amount of 200\,000 particles was kept in the simulations at all times, with blown-out particles being replaced by ones newly spawned along the disk surface.
The resulting dust grain counts are summarised in Table~\ref{tab:particle-counts}.\footnote{Since the entrainment rates for small grains are already very high, radiation pressure \citep[see e.g.][]{Owen-2019, Vinkovic-2021} should not strongly affect our results.
If anything, it would increase the outflow velocity of the dust grains, thus enhancing $\dot{M}_\mathrm{dust}$ by a factor $< 10$.}

\begin{table}
    \centering
    \caption{Statistics for the dust trajectories used to create the density maps: grain size, number of all trajectories modelled, number of fully entrained trajectories thereof, for both transition disk models.}
    \begin{tabular}{crrrr}
        \hline\hline
        & \multicolumn{2}{c}{TD20} & \multicolumn{2}{c}{TD30} \\
        $a_0$ [$\mu$m] & \multicolumn{1}{l}{$N_\mathrm{all}$} & \multicolumn{1}{l}{$N_\mathrm{entrained}$} & \multicolumn{1}{l}{$N_\mathrm{all}$} & \multicolumn{1}{l}{$N_\mathrm{entrained}$} \\ \hline
        0.01 & 3\,897\,202 & 3\,366\,326 & 2\,708\,677 & 2\,466\,946 \\
        0.1 & 3\,782\,149 & 3\,405\,947 & 2\,630\,424 & 2\,427\,226 \\
        0.5 & 3\,324\,900 & 3\,044\,384 & 2\,352\,484 & 2\,147\,716 \\
        1 & 2\,870\,286 & 2\,620\,235 & 1\,997\,824 & 1\,793\,465 \\
        2 & 2\,381\,287 & 2\,125\,436 & 1\,677\,729 & 1\,474\,215 \\
        4 & 1\,760\,818 & 1\,446\,631 & 1\,252\,860 & 1\,038\,191 \\
        8 & 512\,372 & 144\,423 & 604\,993 & 140\,355 \\
        12 & 236\,155 & 8\,640 & 206\,440 & 2\,055 \\ \hline
    \end{tabular}
    \tablefoot{The differing numbers stem from a constant sample size of 200\,000 grains processed simultaneously over similar simulation time spans, with grains being reinserted at a random position along the disk surface once they exit the computational domain.}
    \label{tab:particle-counts}
\end{table}

The resulting trajectories were then used to create dust maps, using the method described in Paper~\citetalias{Franz-2022a} and employing a grid with $\Delta r = 0.5\,$AU for $r < 50\,$AU and $\Delta r = 1\,$AU from there on out, and $\Delta \vartheta = 0.5^\circ$.
The dust grains reach escape velocity within $10^2...10^3\,$yr just like for the PD (see Paper~\citetalias{Franz-2020}), which signifies a much smaller timeframe than the $10^5...10^6\,$yr on which the overall dispersal of the gas happens \citep[see e.g.][]{Mamajek-2009}.
Thus, dust entrainment can still be regarded as a separate process, and using a steady-state gas snapshot should not affect our results.

\subsubsection{Densities}
\label{sec:Methods:dust:rho}

In order to retrieve our dust density estimates for the wind, the dust maps need to be combined with sensible estimates for the dust densities at the wind launching region.
As a basis, we selected 400 grain species, spaced logarithmically in size for $1\,\mathrm{nm} \leq a_0 \leq 1\,\mathrm{mm}$, and drawn from a MRN distribution with $n(a) \propto a^{-3.5}$ \citep{Mathis-1977}.
We then constructed two different setups:
Firstly, we assumed a globally fixed dust-to-gas ratio of 0.01, labelled `fixed'.\footnote{As already pointed out in Paper~\citetalias{Franz-2022a}, real dust-to-gas ratios may be higher \citep[see e.g.][]{Miotello-2017}.}
Secondly, in a setup referred to as `variable' below, we used \texttt{disklab} (Dullemond \& Birnstiel, in prep.) to compute densities accounting for hydrostatic equilibrium in the gas \citep[see e.g.][]{Armitage-2010}, and vertical settling-mixing equilibrium in the dust according to \citet{Fromang-2009}.
The underlying dust-to-gas ratio at the midplane was assumed to be 0.01 like in the `fixed' case.

Preliminary investigations showed that grains $\gtrsim 12.5\,\mu$m were not entrained in the wind; so we discarded all size bins $a_0 > 12.5\,\mu$m from the original 400 species.
The remaining grain species, equalling about 11\% of the total dust mass, were then re-binned into eight bins matching the eight $a_0$ investigated.

In a last step, the dust densities at the disk surface of the `fixed' and `variable' models were used to populate the wind regions via the dust maps created from the trajectories; this yields a total of four setups: `fixed' TD20 (fixTD20), `variable' TD20 (varTD20), `fixed' TD30 (fixTD30), and `variable' TD30 (varTD30).
Each of these `wind' models is complemented by a counterpart with a wind region entirely devoid of dust (`no wind'), to provide a simple means of assessing the effect of the dusty outflow on the observations.
In order to smooth out artefacts resulting from the discrete trajectories and the rather fine grid spacing, we applied a Gaussian filter ($\sigma = 2\,$AU) to the densities in the wind region.

\subsubsection{Mass-loss rates}
\label{sec:Methods:dust:Mdot}

From the HD simulations, we extracted the dust outflow velocities at the domain boundary (i.e. $r \simeq 300\,$AU).
In combination with $\varrho_\mathrm{dust}$, these were then used to compute the dust outflow rate $\dot{M}_\mathrm{dust}$.

\subsection{Radiative transfer}
\label{sec:Methods:images}

\texttt{RadMC-3D} was employed for the radiative-transfer computations.\footnote{\texttt{RadMC-3D}: \href{http://www.ita.uni-heidelberg.de/~dullemond/software/radmc-3d/}{[link]}. Version 2.0 was used for this work.}
In a setup identical to the one used in Paper~\citetalias{Franz-2022a}, we used a stellar temperature of $T_* = 5000\,$K, a logarithmically-spaced wavelength grid with $10^{-1} \leq \lambda \,[\mu\mathrm{m}] \leq 10^4$, and photon counts of $N_\mathrm{phot}^\mathrm{therm} = 10^9$ and $N_\mathrm{phot}^\mathrm{therm} = 10^6$ to create $800 \times 800$ pixel images in scattered and polarised light.
We produced simulated images for $\lambda_\mathrm{obs} \in \lbrace 0.4, 0.7, 1.2, 1.6, 1.8 \rbrace \,\mu$m.
In order to obtain the full Stokes parameters in full anisotropic scattering, we computed the scattering matrix coefficients using \texttt{dsharp\_opac} \citep{Birnstiel-2018}.\footnote{\texttt{dsharp\_opac}: \href{https://github.com/birnstiel/dsharp_opac/}{[link]}.}

For the disk region, we used the opacities from the Disk Substructures at High Angular Resolution Project (DSHARP), composed from the results of \citet{Henning-1996}, \citet{Draine-2003c}, and \citet{Warren-2008}; for the wind region, we employed pure astrosilicate opacities \citep{Draine-2003c}.
This distinction was made because we cannot be sure the wind is cold enough to host ice everywhere.
Furthermore, astrosilicates have a higher albedo, thus slightly enhancing their observability; this matches our intent to investigate best-case scenarios for the observability of the winds.
The impact of choosing this opacity prescription is investigated in App.~\ref{sec:app:opacities}.

\subsubsection{Scattered light}
\label{sec:Methods:images:sca}

The Near Infrared Camera \citep[NIRCam,][]{Rieke-2003, Rieke-2005} of the \textit{James Webb} Space Telescope (JWST) will allow for state-of-the-art scattered-light imaging in the (sub-)$\mu$m wavelength range.
In Paper~\citetalias{Franz-2022a}, we have seen that smaller wavelengths should be most favourable for the detection of a wind signature; so we are mainly interested in the F070W filter ($\lambda_\mathrm{obs} = 0.7\,\mu$m).
Alternatively, in order to use the smallest coronagraph (MASK210R, with a half-transmission radius of $0{\farcs}4$), $\lambda_\mathrm{obs} \gtrsim 1.8\,\mu$m is needed; for our purposes, this means the F182M filter.

\texttt{Mirage} was employed to synthesise the instrument response for JWST NIRCam,\footnote{\texttt{Mirage}: \href{https://mirage-data-simulator.readthedocs.io/}{[link]}. Version 2.1.0 was used for this work.} and the resulting data sets were then post-processed with the \texttt{jwst} pipeline just as an actual observational data set would be.\footnote{\texttt{jwst}: \href{https://jwst-pipeline.readthedocs.io/}{[link]}. Version 1.2.3 was used for this work.}
The coronagraph was simulated by simply applying a transmission mask to the input intensities;\footnote{The transmission curves are available from \href{https://jwst-docs.stsci.edu/near-infrared-camera/nircam-instrumentation/nircam-coronagraphic-occulting-masks-and-lyot-stops}{[here]}.} this should allow for a first-order estimate, although it disregards the peculiarities of the instrument.

\subsubsection{Polarised light}
\label{sec:Methods:images:pol}

We used the same procedure as described in Paper~\citetalias{Franz-2022a} to synthesise an instrument response for the Spectro-Polarimetric High-contrast Exoplanet REsearch \citep[SPHERE,][]{Beuzit-2019} InfraRed Dual-band Imager and Spectrograph (IRDIS) of the Very Large Telescope (VLT).
In particular, we focussed on the $J$- and $H$-bands ($\lambda_\mathrm{obs} \simeq 1.2$ and $1.6\,\mu$m, respectively) in $Q_\phi = Q \cos(2\,\phi) + U \cos(2\,\phi)$, with $\phi = \arctan(x/y)$, and in $P = \sqrt{Q^2+U^2}$.
The noise level was extracted from the comparison between an $H$-band observation of DoAr\,44 \citep{Avenhaus-2018} and a radiative transfer modelling of that object \citep{Casassus-2018}.

Furthermore, we used the predictions for the instrument response to compute the spectral index $\alpha \equiv \alpha_{J,H}$ between $J$- and $H$-bands, both for the `wind' and `no wind' cases ($\alpha_\mathrm{w}$ and $\alpha_\mathrm{nw}$, respectively).
For each model, the difference $\Delta \alpha = \alpha_\mathrm{w} - \alpha_\mathrm{nw}$ was then investigated to check for a possible colour excess caused by the dusty wind.

%%%%%%%%%%%%%%%%%%%%%%%%%%%%%%%%%%%%%%%%%%%%%%%%%%%%%%%%%%%%%%%%%%%%%%%%%%%%%%%%%%%%%%%%%%%%%%%%%%%%

\section{Results}
\label{sec:Results}

\subsection{Trajectories}
\label{sec:Results:dust:traj}

\begin{figure*}[!tb]
    \centering
    \includegraphics[width=0.99\textwidth]{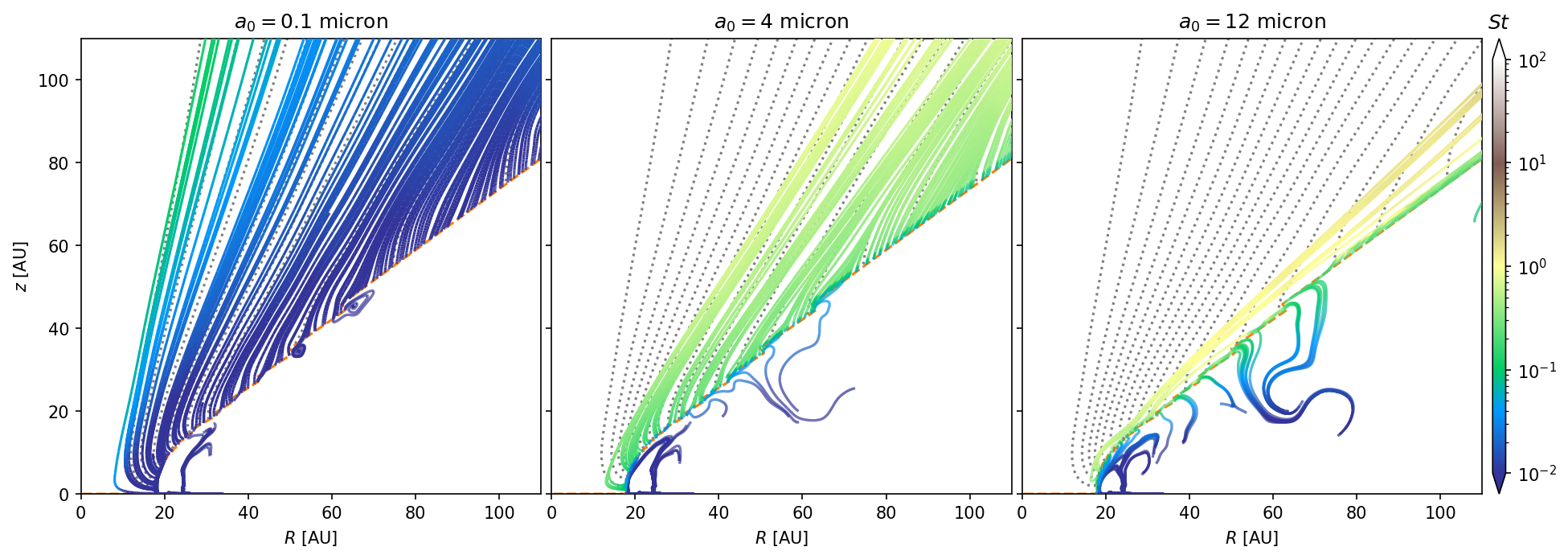}
    \caption{Random selection of dust trajectories in TD20, for three $a_0$; the image has been zoomed in to the inner $110\,\mathrm{AU} \times 110\,\mathrm{AU}$ to give a better view of the region around the outer gap.
    The grains are launched from the disk surface (orange dashed line) and if picked up by the wind, they then follow the gas streamlines (grey dotted lines) more or less closely, depending on their local Stokes number $St$ (colourbar).
    A few of the grains are not entrained, and proceed to move into the disk, following the gas flows there.}
    \label{fig:dust-traj-20}

    \includegraphics[width=0.99\textwidth]{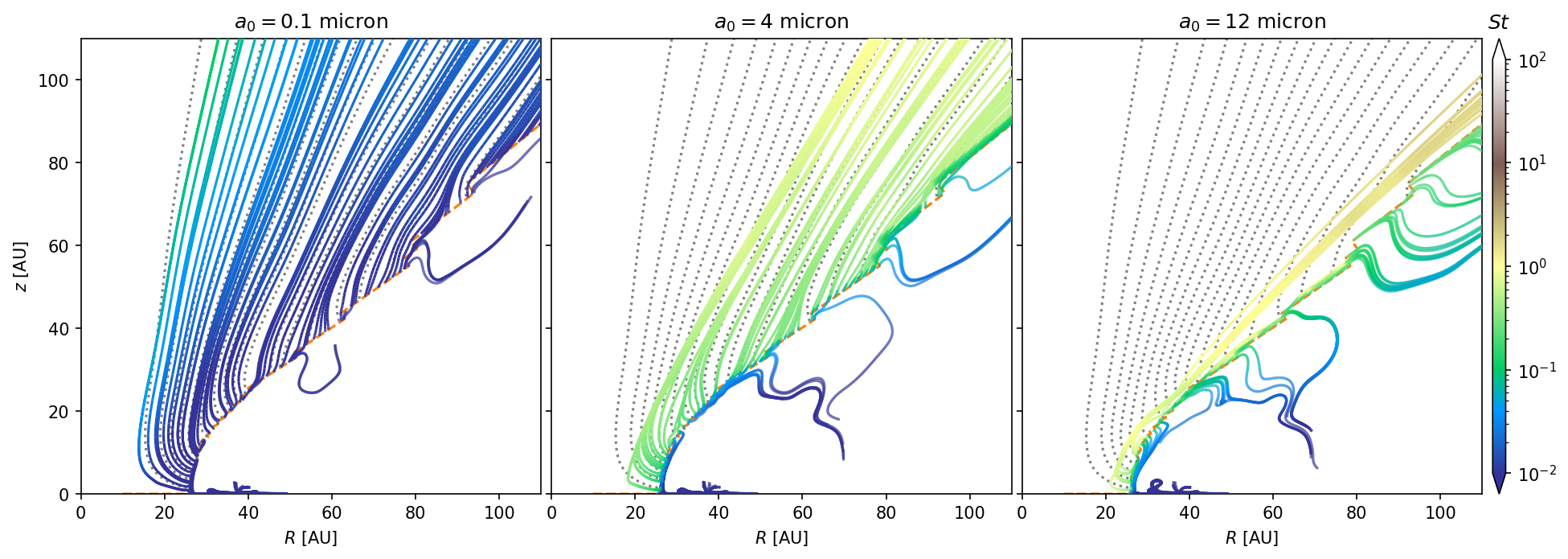}
    \caption{Dust trajectories for various $a_0$ for TD30, all else equal to Fig.~\ref{fig:dust-traj-20}.
    There are no systematic differences in the dust outflow patterns between TD20 and TD30.}
    \label{fig:dust-traj-30}
\end{figure*}

Following the trajectories of grains entrained in the wind, we see that the small grains in our simulation tend to have low Stokes numbers $St$, and thus are well-coupled to the gas phase, resulting in trajectories that hardly differ from the gas streamlines.
Larger grains with higher $St$ decouple much earlier from the gas, and their trajectories depart from the gas streamlines shortly after launch.
Just like for PD, no grains are entrained from $R \gtrsim 140\,$AU, with $R = \sqrt{x^2+y^2}$.
Some randomly selected trajectories are presented in Figs.~\ref{fig:dust-traj-20} (TD20) and \ref{fig:dust-traj-30} (TD30); for each model, we show three grain species to illustrate slow, intermediate and fast decoupling.

The XEUV-driven outflow launching from the outer gap edge is  more vigorous than what we have seen for PD, thus grains up to $a_0 \lesssim 12 \mu$m can be entrained; this however only holds for the strongly curved disk surface region close to the holes.
Their (almost) vertical launch into the wind even from there, and for all grain sizes, corroborates the assumption of a vertical launch from a non-angled surface made by \citet{Clarke-2016} for a semi-analytical EUV-only gas model.
These have since been refined to include dust by \citet{Hutchison-2021}, and extended to allow for various launching angles by \citet{Sellek-2021}.
The qualitative agreement between our numerical results and the semi-analytical work above would suggest that in theory costly hydrodynamical simulations as presented in this work could be replaced by semi-analytical formulations.
In practice however, as \citet{Booth-2021a} have shown, differing assumptions as to the vertical delivery of the dust to the wind launching region will lead to significantly different density estimates for the wind region.

\subsection{Dust densities}
\label{sec:Results:dust:rho}

\begin{figure*}[!t]
    \centering
    \includegraphics[width=0.99\textwidth]{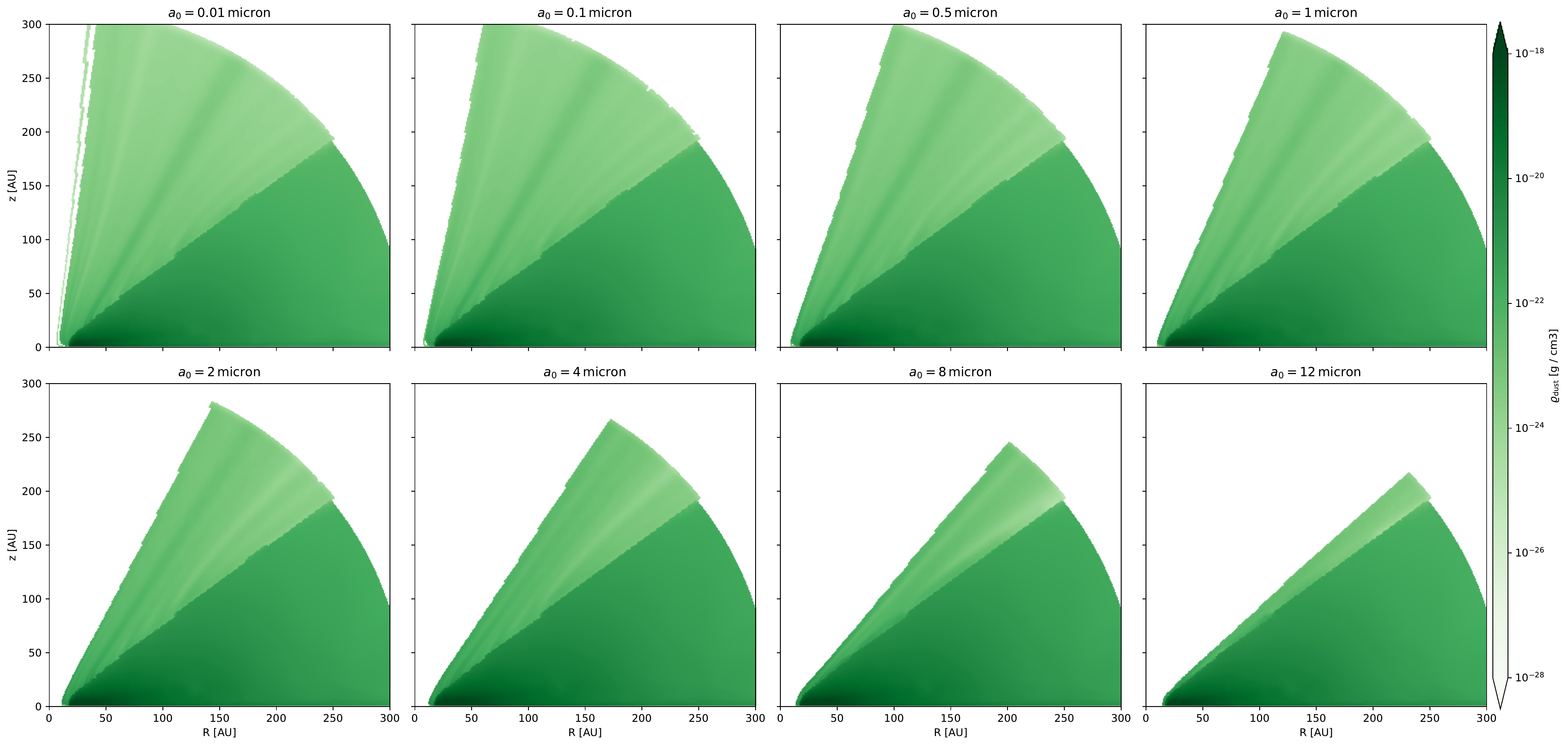}
    \caption{Dust densities for the XEUV-irradiated fixTD20 disk in $(R, z)$.
    The presence of small dust grains everywhere along the disk surface leads to a mostly smooth distribution of dust densities in the wind.}
    \label{fig:dust-rho-20-fix}

    \includegraphics[width=0.99\textwidth]{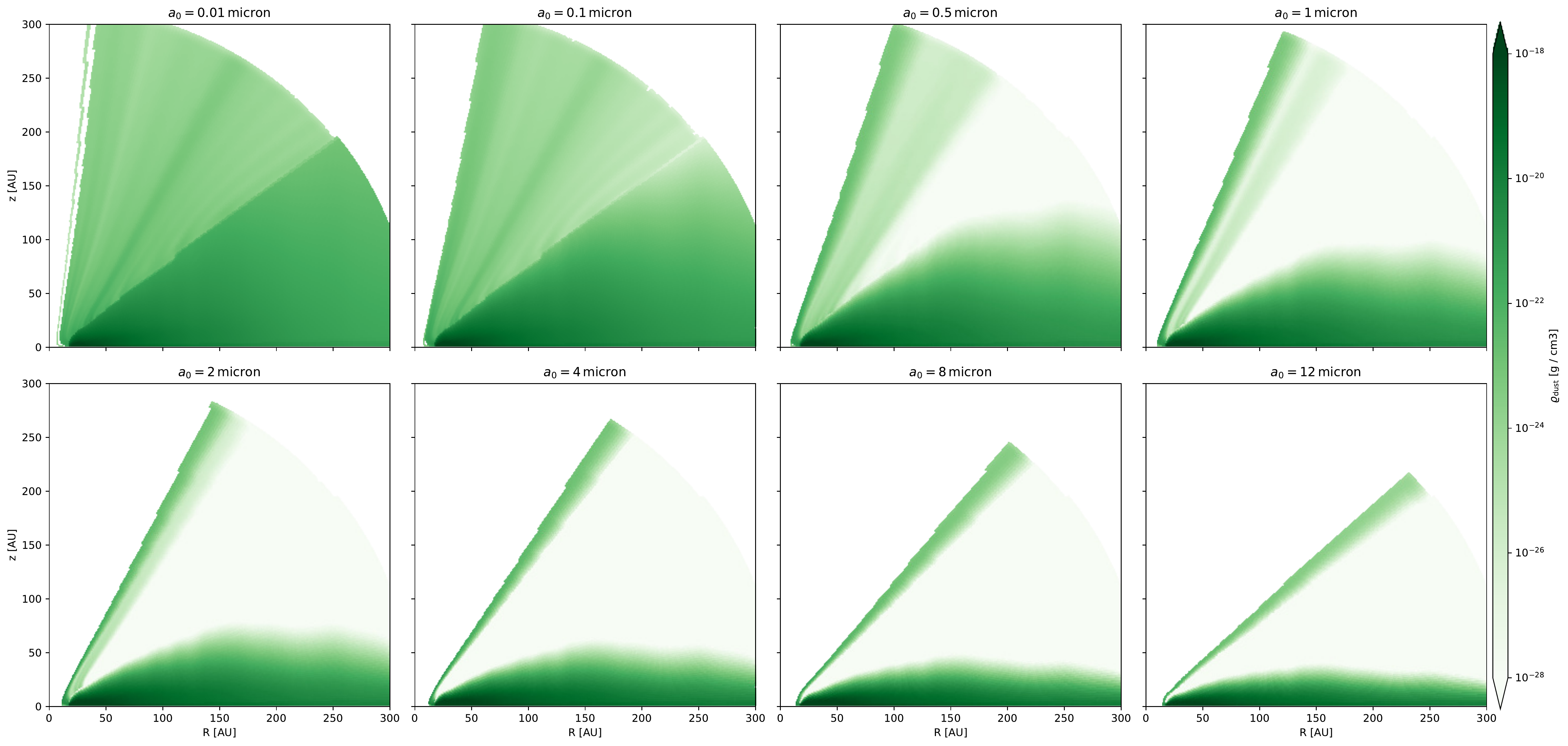}
    \caption{varTD20 counterpart of Fig.~\ref{fig:dust-rho-20-fix}, using an identical colourbar.
    Dust settling leads to an outflow which, for $a_0 \gtrsim 0.5\,\mu$m, is mainly fueled from the outer gap edge where the wind interacts with the disk at low $z$.}
    \label{fig:dust-rho-20-var}
\end{figure*}

\begin{figure*}[!t]
    \centering
    \includegraphics[width=0.99\textwidth]{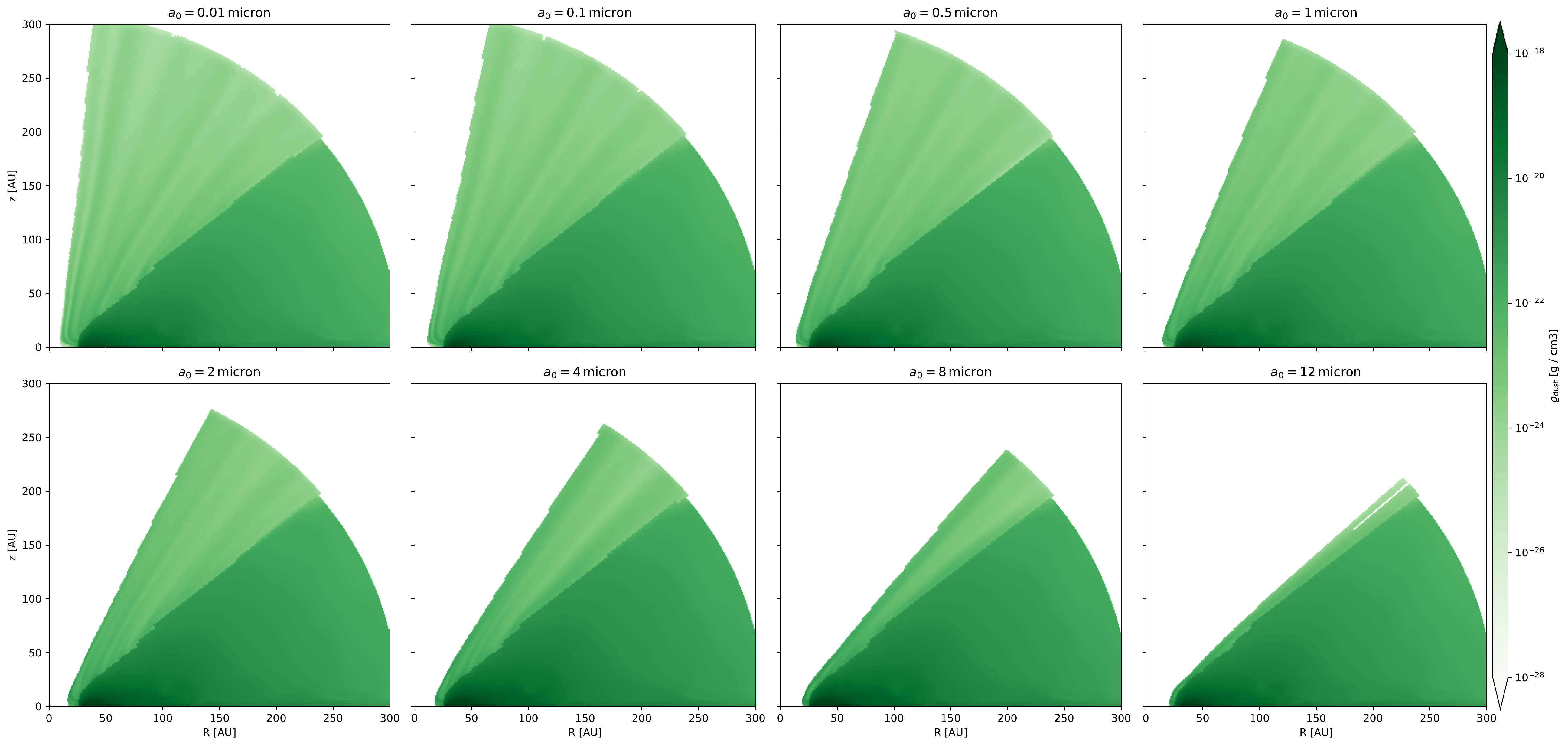}
    \caption{Dust densities for fixTD30, all else equal to Fig.~\ref{fig:dust-rho-20-fix}.
    Apart from the larger inner hole, the differences to the other `fixed' model (Fig.~\ref{fig:dust-rho-20-fix}) are minor.}
    \label{fig:dust-rho-30-fix}

    \includegraphics[width=0.99\textwidth]{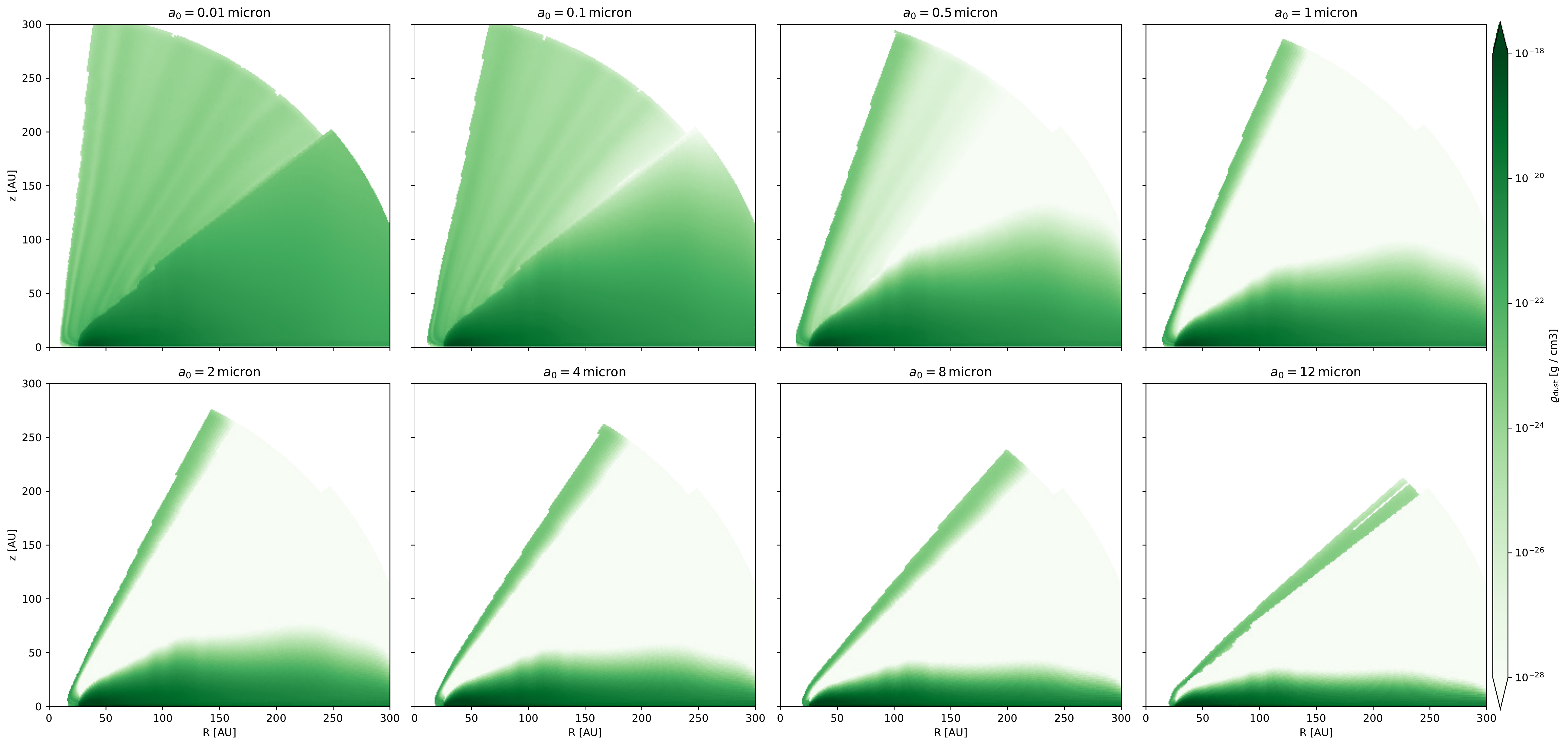}
    \caption{Dust densities for varTD30, all else equal to Fig.~\ref{fig:dust-rho-30-fix}.
    As for the `fixed' models, the differences between the varTD20 (Fig.~\ref{fig:dust-rho-20-var}) and varTD30 disks in terms of dust content in the wind are minor.}
    \label{fig:dust-rho-30-var}
\end{figure*}

The dust densities computed for our various models are shown in Figs.~\ref{fig:dust-rho-20-fix} (fixTD20), \ref{fig:dust-rho-20-var} (varTD20), \ref{fig:dust-rho-30-fix} (fixTD30) and \ref{fig:dust-rho-30-var} (varTD30).
The choice of dust settling prescription has a larger impact on the morphology of the dusty wind than the $\approx 10\,$AU difference in hole size between TD20 and TD30.

For the `fixed' setups (Figs.~\ref{fig:dust-rho-20-fix} and \ref{fig:dust-rho-30-fix}), the wind region is continuously populated by dust grains between the disk surface and a maximum reachable scale height (see Paper~\citetalias{Franz-2020}), even though there are some outflow channels with slightly enhanced densities.
This stands in stark contrast to the `variable' models (Figs.~\ref{fig:dust-rho-20-var} and \ref{fig:dust-rho-30-var}).
In the latter, dust with $a_0>0.1\,\mu$m only populates the wind regions in significant quantities if it has been launched from close to the edge of the hole, that is from low $z$; this leads to quite distinct outflow channels, which become narrower for larger $a_0$, and are close to the maximum scale height $\left. \max(z) \right|_R$ the dust reaches in the wind (taken individually for each $a_0$, compare Paper~\citetalias{Franz-2020}).
For $a_0 \lesssim 4\,\mu$m, the densities in the dominant outflow channels of the `variable' models are very similar to those found in these regions in the `fixed' models.

This concentrated outflow pattern stands in clear contrast to the much smoother dust density maps we have seen for PD.
The smallest grains ($a_0=0.01\,\mu$m), which are (almost) fully hydrodynamically coupled with the gas, still occupy as much of the wind region as in the `fixed' cases, which serves as a sanity check.
Contrary to PD, even the largest grains show clear non-zero wind densities.

Several recent works have investigated the maximum entrainable $a_0$ for MHD \citep{Miyake-2016, Giacalone-2019}, EUV \citep{Hutchison-2016c, Hutchison-2016b, Hutchison-2021}, and XEUV winds \citep[][and Paper~\citetalias{Franz-2020}]{Booth-2021a}.
However, as we can see from our `variable' models (Figs.~\ref{fig:dust-rho-20-var} and \ref{fig:dust-rho-30-var}), the gas densities cannot simply be converted into dust densities via a global ratio when invoking vertical settling.
So as laid out in Paper~\citetalias{Franz-2022a} \citep[and also stated by many other authors, e.g.][]{deBoer-2017, Villenave-2020}, observational constraints on the strength of the vertical settling in protoplanetary disks are needed for more accurate modelling.

\subsection{Mass-loss rates}
\label{sec:Results:dust:Mdot}

For PD, we calculated $\dot{M}_\mathrm{dust} / \dot{M}_\mathrm{gas} \lesssim 1.1 \cdot 10^{-3}$ and $3.2 \cdot 10^{-4}$ for the `fixed' and `variable' cases, respectively (Paper~\citetalias{Franz-2022a}).
The upper limit corresponds to the dust-to-gas ratio assumed (0.01) multiplied by the mass fraction of the entrainable grains in relation to the total dust population ($\approx 11$\%).

The XEUV-driven mass-loss rates for the transition disks modelled here are listed in Table~\ref{tab:massloss}.
Within the `fixed' and `variable' setups, $\dot{M}_\mathrm{dust} / \dot{M}_\mathrm{gas}$ has increased for the disks with an inner hole.
This stands to reason because the inner hole allows the photoevaporative wind to directly penetrate to, and thus entrain material from, regions close to the disk midplane.

\begin{table*}[!bt]
    \centering
    \caption{Gas and dust mass-loss rates $\dot{M}_\mathrm{gas}$ and $\dot{M}_\mathrm{dust}$ for the models, in units of [M$_\odot$/yr].}
    \def\arraystretch{1.2}
    \begin{tabular}{c*{4}{c}}
        \hline \hline
        & \multicolumn{4}{c}{$\dot{M}_\mathrm{dust}$ [M$_\odot$/yr]} \\
        $a_0$ $[\mu\mathrm{m}]$ & fixTD20 & varTD20 & fixTD30 & varTD30 \\ \hline
        0.01 & $7.6 \cdot 10^{-12}$ & $6.6 \cdot 10^{-12}$ & $6.4 \cdot 10^{-12}$ & $5.0 \cdot 10^{-12}$ \\
        0.1 & $1.1 \cdot 10^{-11}$ & $3.1 \cdot 10^{-12}$ & $9.1 \cdot 10^{-12}$ & $2.2 \cdot 10^{-12}$ \\
        0.5 & $9.8 \cdot 10^{-12}$ & $1.8 \cdot 10^{-12}$ & $7.2 \cdot 10^{-12}$ & $9.9 \cdot 10^{-13}$ \\
        1 & $9.8 \cdot 10^{-12}$ & $1.7 \cdot 10^{-12}$ & $6.9 \cdot 10^{-12}$ & $1.0 \cdot 10^{-12}$ \\
        2 & $1.0 \cdot 10^{-11}$ & $1.7 \cdot 10^{-12}$ & $7.2 \cdot 10^{-12}$ & $9.1 \cdot 10^{-13}$ \\
        4 & $9.3 \cdot 10^{-12}$ & $1.4 \cdot 10^{-12}$ & $6.1 \cdot 10^{-12}$ & $6.1 \cdot 10^{-13}$ \\
        8 & $3.2 \cdot 10^{-12}$ & $3.9 \cdot 10^{-13}$ & $2.1 \cdot 10^{-12}$ & $2.4 \cdot 10^{-13}$ \\
        12 & $2.1 \cdot 10^{-13}$ & $2.3 \cdot 10^{-14}$ & $8.9 \cdot 10^{-14}$ & $2.8 \cdot 10^{-14}$ \\ \hline
        (sum) & $6.2 \cdot 10^{-11}$ & $1.7 \cdot 10^{-11}$ & $4.5 \cdot 10^{-11}$ & $1.1 \cdot 10^{-11}$ \\ \hline
        $\dot{M}_\mathrm{dust} / \dot{M}_\mathrm{gas}$ & $2.0 \cdot 10^{-3}$ & $5.5 \cdot 10^{-4}$ & $1.6 \cdot 10^{-3}$ & $3.8 \cdot 10^{-4}$ \\ \hline
    \end{tabular}
    %\tablefoot{}
    \label{tab:massloss}
\end{table*}

%The fact that $\dot{M}_\mathrm{dust} / \dot{M}_\mathrm{gas} > 1.1 \cdot 10^{-3}$ is clearly exceeded despite being the theoretical maximum may indicate that limiting the disk surface to 1D may negatively affect the accuracy of our model especially in regards to where material is entrained from at the outer gap edge.
%In addition, our basic assumption that the dust distribution directly follows the gas distribution (especially for the `fixed' model) is an approximation \citep[see e.g.][]{Testi-2014}.
%However, an enhanced $\dot{M}_\mathrm{dust}$ is not detrimental to our goal of presenting a best-case scenario for the observability of dusty photoevaporation-driven outflows.
%Furthermore, it (over)compensates for the -- probably unphysical -- decrease in $\dot{M}_\mathrm{dust}$ compared to PD \citep{Picogna-2019}.

While $\dot{M}_\mathrm{gas}$ decreases only slightly from TD20 to TD30, there is a clear decline in $\dot{M}_\mathrm{dust}$ between the models.
This is most likely due to the gas surface mass-loss rate $\dot{\Sigma}_\mathrm{gas}$ being higher at the gap edge for TD20, leading to an overall stronger dusty outflow.

All of the recorded cumulative values for $\dot{M}_\mathrm{dust}$ are above their counterparts for PD, which were $\dot{M}_\mathrm{dust}^\text{`fix'} \simeq 4.1 \cdot 10^{-11} \,\mathrm{M_\odot/yr}$ and $\dot{M}_\mathrm{dust}^\text{`var'} \simeq 1.2 \cdot 10^{-11} \,\mathrm{M_\odot/yr}$ alongside $\dot{M}_\mathrm{gas} \simeq 3.7 \cdot 10^{-8} \,\mathrm{M_\odot/yr}$.
With respect to the dust, this matches the general consensus that photoevaporative mass loss is enhanced in transition disks compared to primordial objects \citep[see e.g.][]{Clarke-2001, Ercolano-2017b}; it additionally coincides with a very rough time estimate for when the gas and dust mass of the disk become similar:
If we assume $M_\mathrm{disk}^\mathrm{(dust)} = 0.01 \, M_\mathrm{disk}^\mathrm{(gas)}$, and $\dot{M}_\mathrm{dust}$ to be constant, we find $M_\mathrm{disk}^\mathrm{(dust)} \gtrsim \, M_\mathrm{disk}^\mathrm{(gas)}$ for $t \gtrsim 0.13\,$Myr (TD20) or 0.11\,Myr (TD30); for PD, this value was $t \gtrsim 0.15\,$Myr.
Considering that our values for $\dot{M}_\mathrm{dust}$ may be overestimates as laid out above, these values may be not entirely accurate, but do show that an enhancement of the XEUV-driven mass loss is quite likely once our model has reached its transitional phase.

\subsection{Scattered-light imaging}
\label{sec:Results:sca}

\subsubsection{Radiative transfer}
\label{sec:Results:sca:radmc}

\begin{figure*}[!t]
    \centering
    \includegraphics[width=0.99\textwidth]{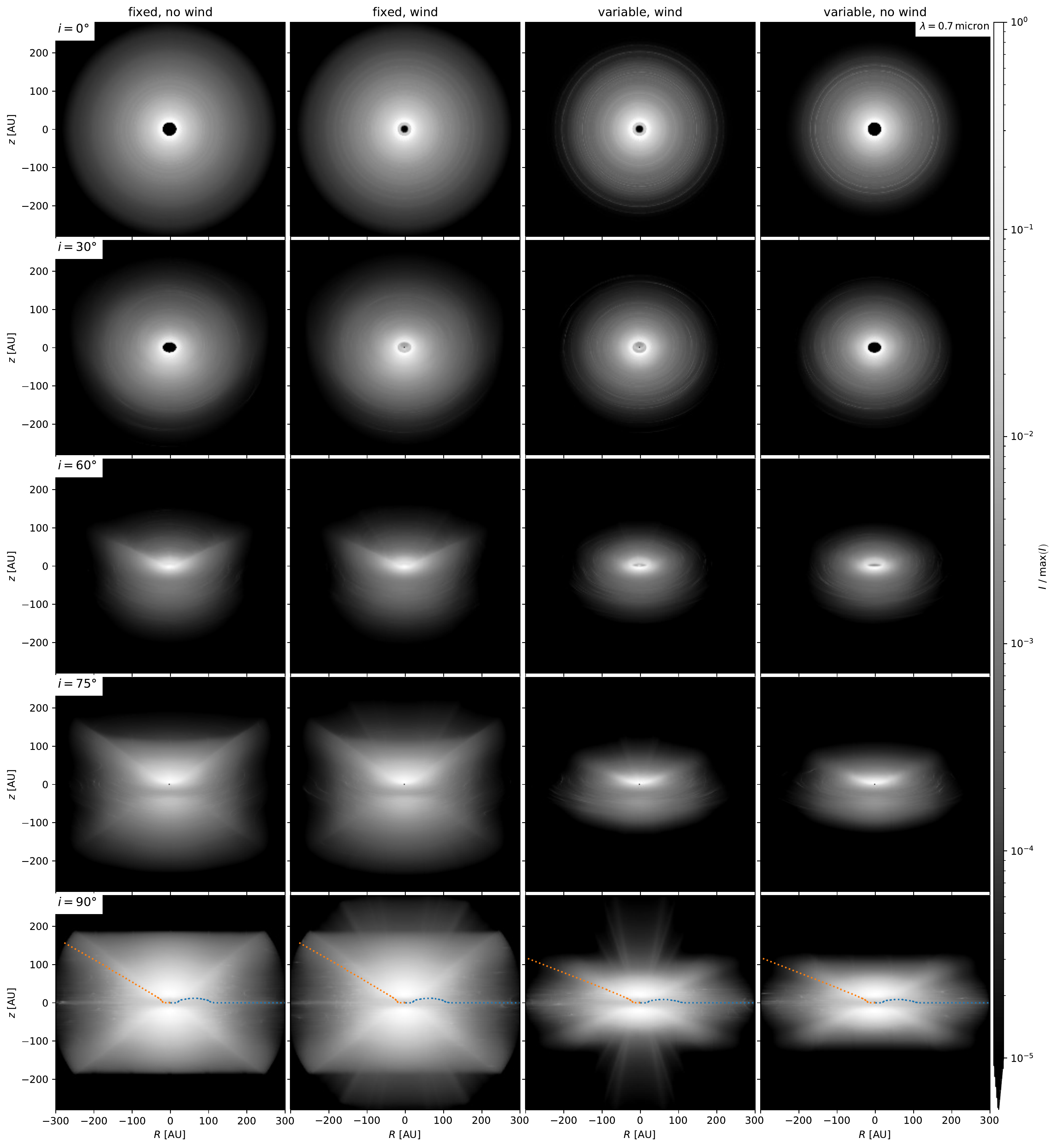}
    \caption{Scattered-light intensities for $\lambda_\mathrm{obs} = 0.7\,\mu$m for TD20; different models in different \textit{columns}, and different inclinations in different \textit{rows}.
    An artificial coronagraph of $r = 1\,$AU is used to mask out the direct stellar signal.
    The $\max(I)$ for the scaling of the colourbar is taken for each image individually (after application of the coronagraphic mask).
    The orange and blue lines in the ($i = 90^\circ$)-row represent the ($\tau = 1$)-surfaces for an observer at $r = 0$ and $z = \infty$, respectively.
    At low $i$, the dusty wind obscures the cavity; at higher $i$, it produces a cone-like feature around the $z$-axis.}
    \label{fig:radmc-sci-20-0.7}
\end{figure*}

\begin{figure*}
    \centering
    \includegraphics[width=0.99\textwidth]{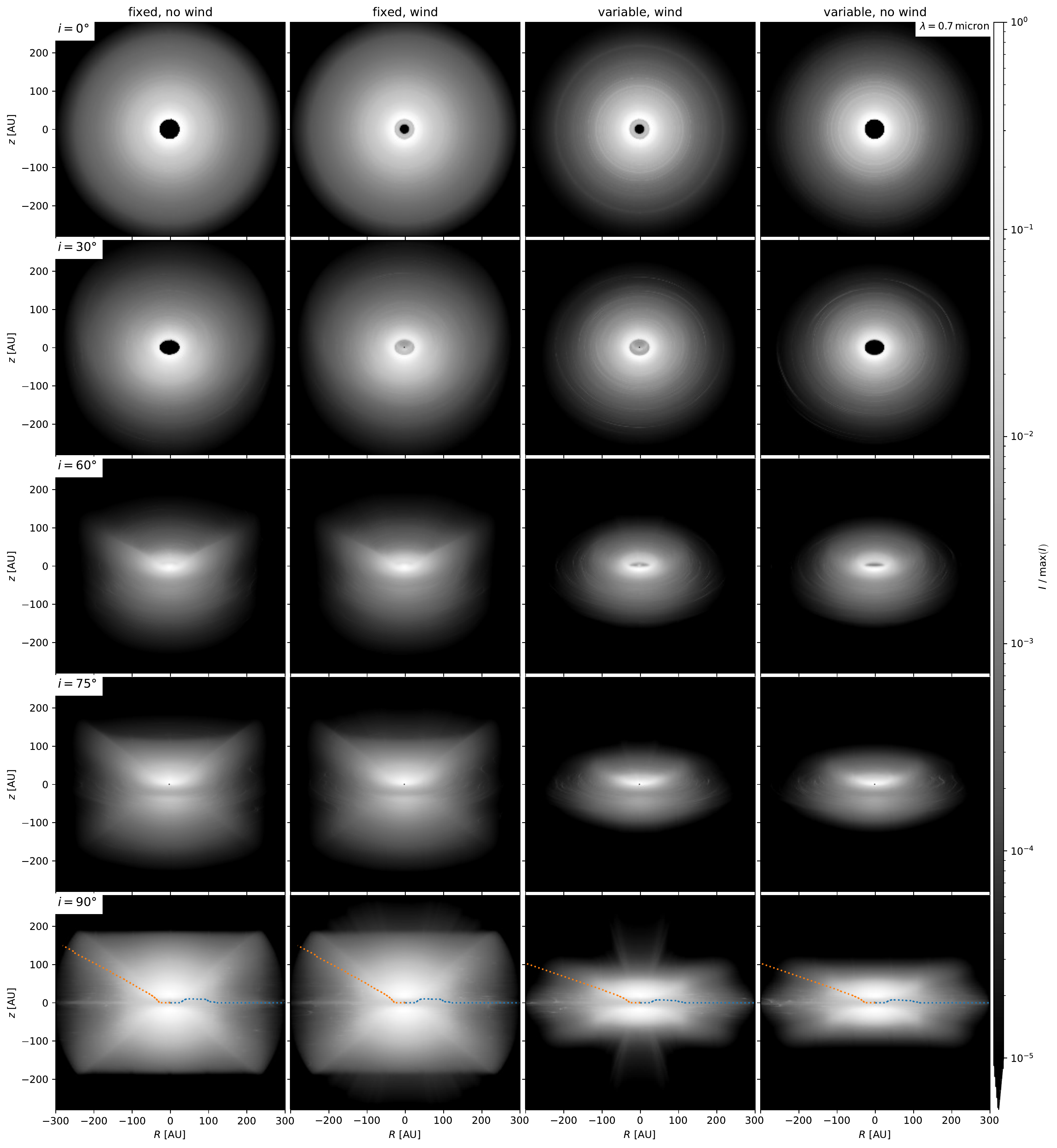}
    \caption{Scattered-light intensities for $\lambda_\mathrm{obs} = 0.7\,\mu$m for TD30, all else equal to Fig.~\ref{fig:radmc-sci-20-0.7}.
    The wind features are similar to those of TD20, but less distinct at high $i$.}
    \label{fig:radmc-sci-30-0.7}
\end{figure*}

We performed radiative transfer calculations with \texttt{RadMC-3D}.\footnote{In order to keep the manuscript concise, we did not include the results for all $\lambda_\mathrm{obs}$ here. Additional plots are however provided in App.~\ref{sec:app:images}.}
The resulting images at $\lambda_\mathrm{obs} = 0.7\,\mu$m, which is the shortest wavelength accessible with JWST, are shown in Figs.~\ref{fig:radmc-sci-20-0.7} (TD20) and \ref{fig:radmc-sci-30-0.7} (TD30). 
To avoid saturating the central region and to enhance the visibility of the fainter outer disk and wind, we applied an artificial coronagraph of $r = 1\,$AU to the images in post-processing to block out the direct stellar signal.
This gives a contrast boost of $\lesssim \lbrace 3.6, 3.2, 2.6, 0.8, 0.0 \rbrace \,$dex (TD20) or $\lesssim \lbrace 4.0, 3.6, 2.9, 1.9, 0.0 \rbrace \,$dex (TD30) at $i = \lbrace 0, 30, 60, 75, 90 \rbrace^\circ$.

At low inclinations ($i \leq 30^\circ$ for the cases investigated here), the `wind' and `no wind' models distinctly differ in the diameter of their inner holes.
For TD20 (TD30) and at $i = 0^\circ$, the scattered light from the dust in the wind extends inwards to $r \gtrsim 7\,$AU (11\,AU), in contrast to the $r \gtrsim 18\,$AU (26\,AU) of the `no wind' models; this corresponds to the regions which are populated by dust grains because the dust entrained from the outer gap edge moves inwards before being blown out of the domain (see Sect.~\ref{sec:Results:dust:traj}).
The apparent difference in hole size is even more pronounced at $i = 30^\circ$, with the diffuse radiation from the dust in the wind covering the full inner hole.
At $i \gtrsim 60^\circ$, the effect disappears because a non-negligible amount of dust is located along the line of sight.
Instead, we find a more distinct wind signature.
As in Paper~\citetalias{Franz-2022a}, when comparing the `wind' and `no wind' images for $i \geq 60^\circ$, we find an outflow pattern that appears like a cone (or chimney) around the polar ($z$-) axis of the disk.
It is quite faint at $i=60^\circ$ and rather pronounced at $i = 90^\circ$.

For the `fixed' model, the wind cone is comparably wide and evenly illuminated.
By contrast, the chimney of the `variable' model is much more condensed towards $\left. \max(z) \right|_R$; this corresponds to the underlying dust distributions (see Sect.~\ref{sec:Results:dust:rho}).
While the amount of dust around $\left. \max(z) \right|_R$ is similar between the `fixed' and `variable' models, there is little dust elsewhere in the wind region for the latter setup; this leads to a more collimated appearance of the cone.
Furthermore, due to the lower vertical extent of the dusty disk in the `variable' models, the dust quantities in the wind exceed those of the disk at lower $z$ and thus $r$; hence the outflowing material is illuminated more strongly by the star.
The resulting steep chimney feature could potentially be used to identify a scenario in which the dust is entrained primarily from the inner gap edge.

Despite the rather small $\lambda_\mathrm{obs}$ and accordingly large fractions of similarly-sized dust grains entrained in the wind (see Table~\ref{tab:particle-counts}), the wind region is optically thin.
Looking at the optical depth $\tau$, the ($\tau = 1$)-surfaces for the corresponding `wind' and `no wind' models are almost identical for an observer placed both at $r = 0$ and $z = \infty$; this was already the case for PD (Paper~\citetalias{Franz-2022a}).

With respect to PD, the relative brightness of the wind features in transition disks is distinctly higher.
At $\lambda_\mathrm{obs} = 0.4\,\mu$m, we found wind intensities $I / I_{\max} \lesssim 10^{-3.5}$, $I_{\max} \equiv \max(I)$, for PD; the cones of the transition disks emerge already at $I / I_{\max} \lesssim 10^{-2}$ (TD20) and $I / I_{\max} \lesssim 10^{-3}$ (TD30).
At $\lambda_\mathrm{obs} = 0.7\,\mu$m, the values are $I / I_{\max} \lesssim 10^{-3.5}$ (TD20) and $I / I_{\max} \lesssim 10^{-4.5}$ (TD30) for the most luminous features of the `variable' disks, underlining the importance of observations at short $\lambda_\mathrm{obs}$.

\begin{figure*}
    \centering
    \includegraphics[width=0.99\textwidth]{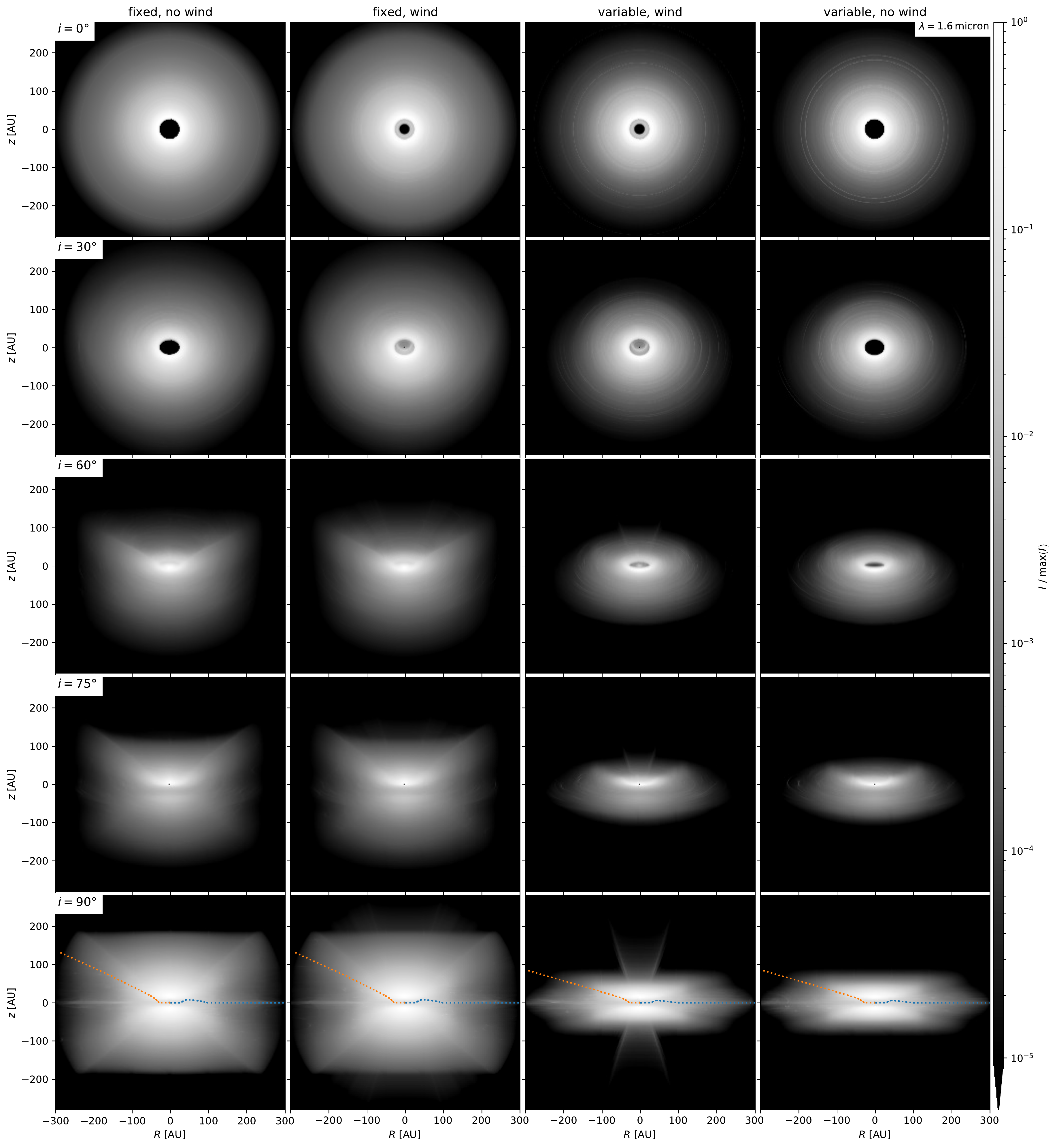}
    \caption{Scattered-light intensities for $\lambda_\mathrm{obs} = 1.6\,\mu$m for TD20, all else equal to Fig.~\ref{fig:radmc-sci-20-0.7}.
    Vertical settling of the dust in the disk enhances the relative strength of the cone-like outflow signature.}
    \label{fig:radmc-sci-20-1.6}
\end{figure*}

\begin{figure*}
    \centering
    \includegraphics[width=0.99\textwidth]{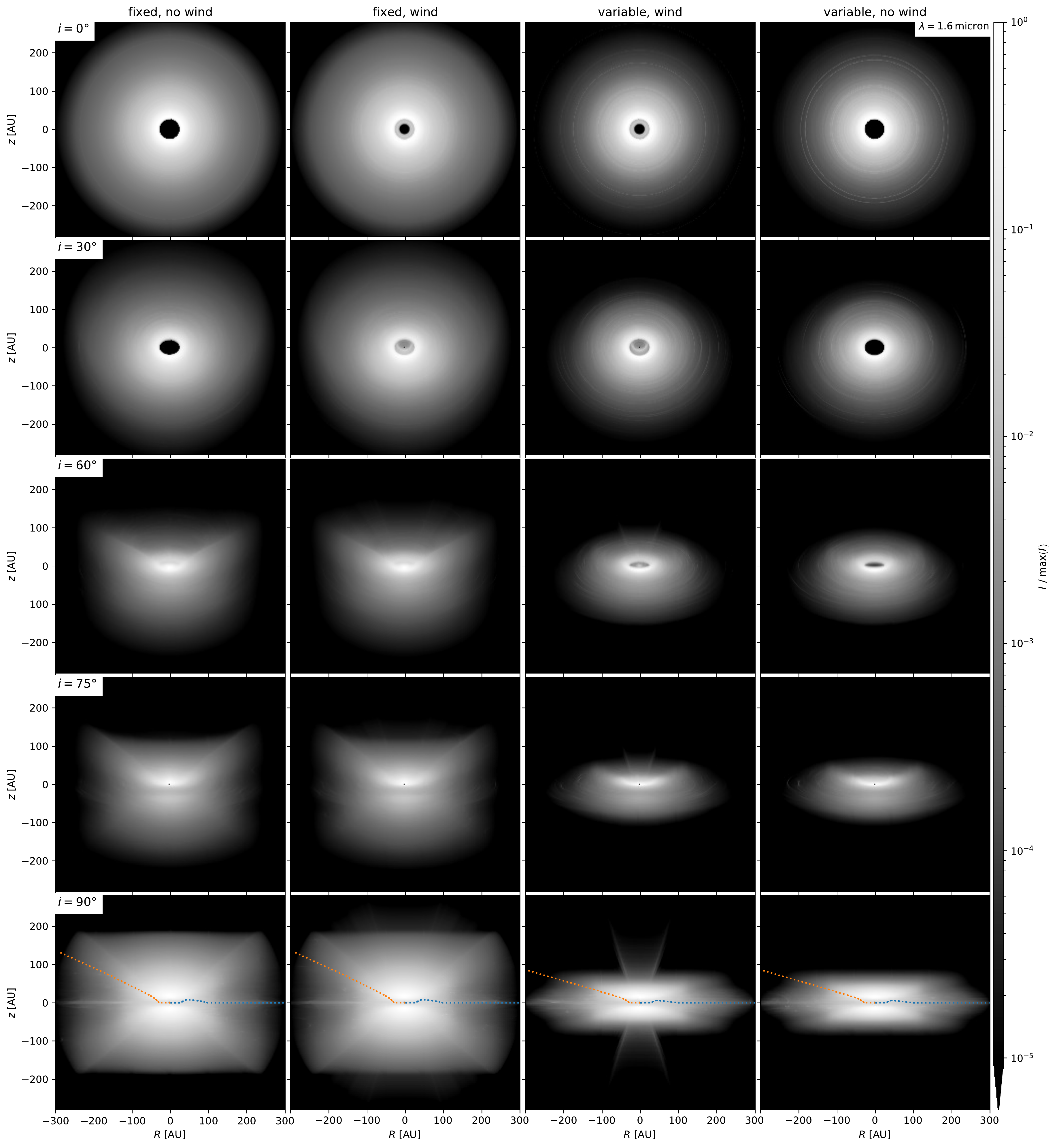}
    \caption{Scattered-light intensities for $\lambda_\mathrm{obs} = 1.6\,\mu$m for TD30, all else equal to Fig.~\ref{fig:radmc-sci-20-0.7}.
    At this wavelength, the differences to TD20 (Fig.~\ref{fig:radmc-sci-20-1.6}) are minor.}
    \label{fig:radmc-sci-30-1.6}
\end{figure*}

Figs.~\ref{fig:radmc-sci-20-1.6} and \ref{fig:radmc-sci-30-1.6} show the \texttt{RadMC-3D} scattered-light intensities for $\lambda_\mathrm{obs} = 1.6\,\mu$m.
Here, the artificial coronagraph grants a contrast boost of $\lesssim \lbrace 3.5, 3.1, 2.7, 1.8, 0.0 \rbrace \,$dex (TD20) or $\lesssim \lbrace 4.0, 3.5, 3.0, 2.4, 0.0 \rbrace \,$dex (TD30).
The main difference compared to $\lambda_\mathrm{obs} = 0.7\,\mu$m is that the wind cone disappears for the `fixed' setup.
By contrast, it persists for the `variable' setup which, due to the vertical settling of the dust grains, has a much lower dust disk scale height, making the chimney more distinct at similar $z$.
The relative intensities are $I / I_{\max} \lesssim 10^{-3.5}$ (TD20) and $I / I_{\max} \lesssim 10^{-4}$ (TD30), versus $I / I_{\max} \lesssim 10^{-4.5}$ for PD.

\subsubsection{Synthetic observations for JWST NIRCam}
\label{sec:Results:sca:jwst}

As in Paper~\citetalias{Franz-2022a}, we encountered clear overexposure issues when synthesising the instrument response for JWST NIRCam via \texttt{Mirage}, polluting the image quite far out.
Again, these can be circumvented by employing the MASK210R coronagraph, but that means that the smallest-wavelength filter usable is F182M, not F070W.

\begin{figure*}
    \centering
    \includegraphics[width=0.99\textwidth]{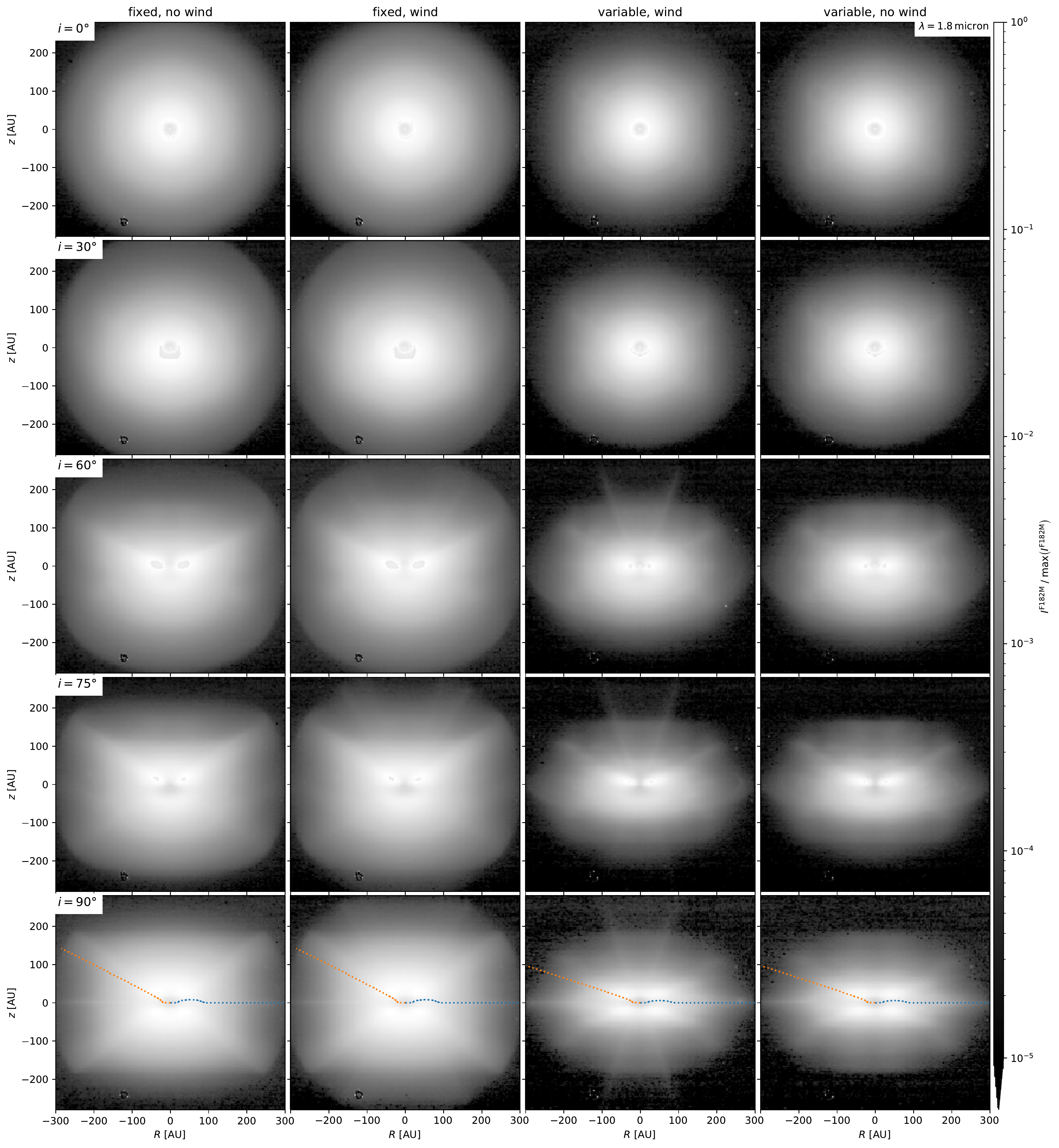}
    \caption{Synthesised observations of TD20 with JWST NIRCam's F182M filter, assuming the MASK210R coronagraph and MEDIUM2 readout pattern.
    The coloured lines indicate the $(\tau=1)$-surfaces as in Figs.~\ref{fig:radmc-sci-20-0.7}--\ref{fig:radmc-sci-30-1.6}.
    At higher $i$, the cone-shaped outflow feature is visible for both the `fixed' and `variable' models.}
    \label{fig:jwst-20}
\end{figure*}

\begin{figure*}
    \centering
    \includegraphics[width=0.99\textwidth]{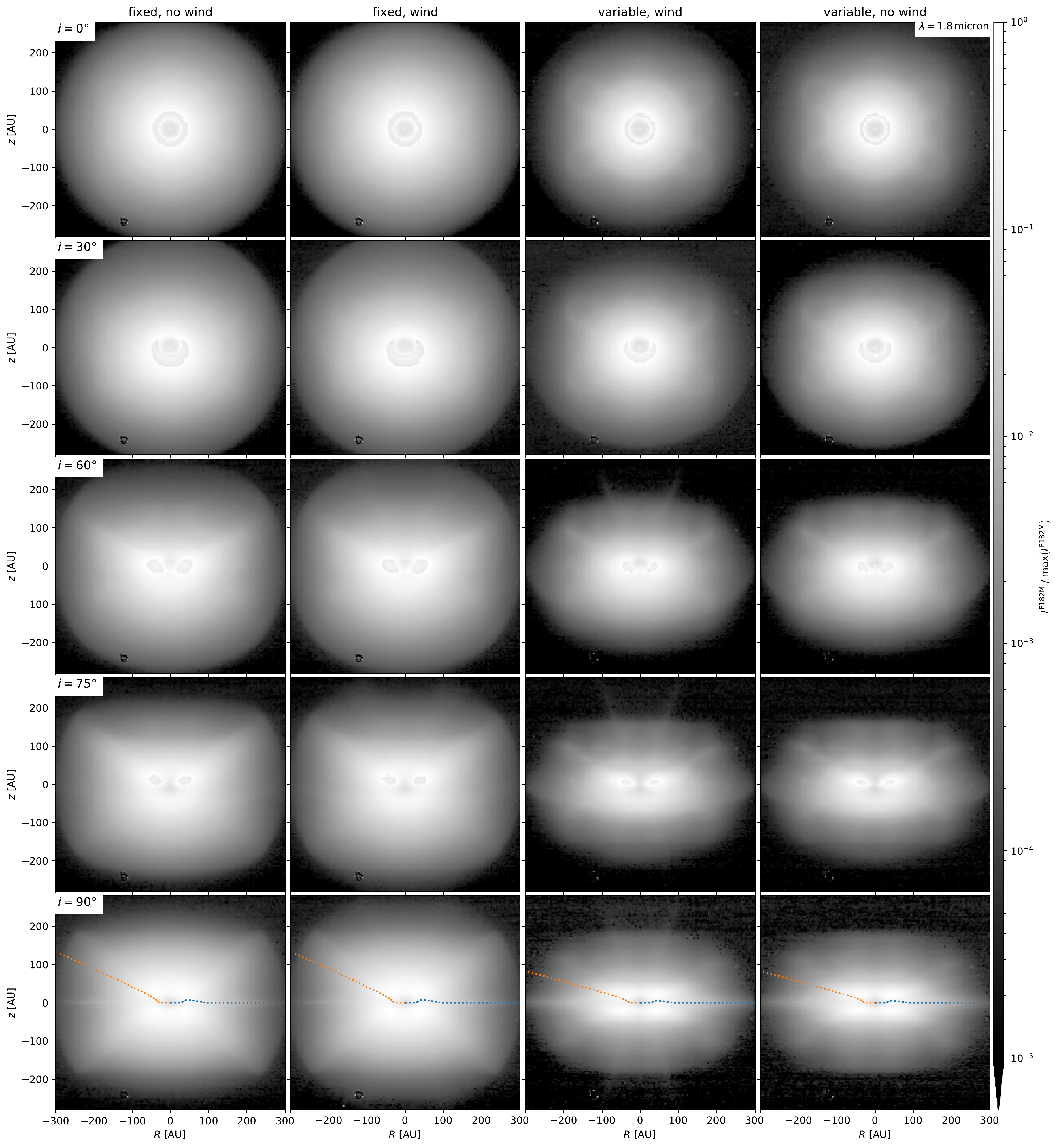}
    \caption{Synthesised observations of TD30 with JWST NIRCam's F182M filter, assuming the MASK210R coronagraph and MEDIUM2 redout pattern; all else equal to Fig.~\ref{fig:jwst-20}.
    The relative intensity of the wind signature is slightly smaller than for TD20.}
    \label{fig:jwst-30}
\end{figure*}

The synthesised coronagraphic images are shown in Figs.~\ref{fig:jwst-20} (TD20) and \ref{fig:jwst-30} (TD30).
For these images, the MEDIUM2 readout pattern was used; shorter science durations result in more noise obscuring potential features, while longer times lead to a more pronounced overexposure of the central regions, which may bleed out into high-$r$ areas.
Despite the transmission mask, the innermost region still exhibits high brightness; this is due to the point-spread function of the instrument.\footnote{For illustrations of the instrument PSF, see \href{https://jwst-docs.stsci.edu/jwst-near-infrared-camera/nircam-predicted-performance/nircam-point-spread-functions}{[link]}.}
The coronagraph diameter of $0{\farcs}8$ ($\simeq 80\,$AU at the assumed distance of 100\,pc) renders a differentiation of the `wind' and `no wind' models by their inner hole sizes unfeasible.

As in the direct \texttt{RadMC-3D} results, the TD20 models show a more prominent wind signature than the TD30 ones.
The `fixed' setup produces a somewhat fuzzy wind signature at higher inclinations, that is $i \gtrsim 60^\circ$ for fixTD20 and $i \gtrsim 75^\circ$ for fixTD30.
Depending on the actual outer radius of the dusty disk (here assumed to be $r \simeq 300\,$AU), the chimney structure caused by the wind may be more or less visible in reality; looking at the fully edge-on disks, it is furthermore questionable whether a clear distinction between a `wind' case and a slightly puffed-up disk could be made from a single observational image.

The flatter disk structure of the `variable' models allows their dusty XEUV outflow pattern to emerge more clearly and already at lower inclinations than for their `fixed' counterparts.
A faint, non-distinct vertical bump already appears at $i=30^\circ$ for varTD20, and a more pronounced cone feature at $i \gtrsim 60^\circ$ for both varTD20 and varTD30.

\subsection{Polarised-light images}
\label{sec:Results:pol}

SPHERE IRDIS's $J$-band corresponds to $\lambda_\mathrm{obs} \simeq 1.2\,\mu$m, thus we chose to present our results for this wavelength here.
As smaller dust grains are more likely to be entrained in the XEUV outflow, $H$-band observations (i.e. $\lambda_\mathrm{obs} \simeq 1.6\,\mu$m) exhibit less distinct wind features (see App.~\ref{sec:app:images}).

\subsubsection{Radiative transfer}
\label{sec:Results:pol:radmc}

\begin{figure*}[!t]
    \centering
    \includegraphics[width=0.99\textwidth]{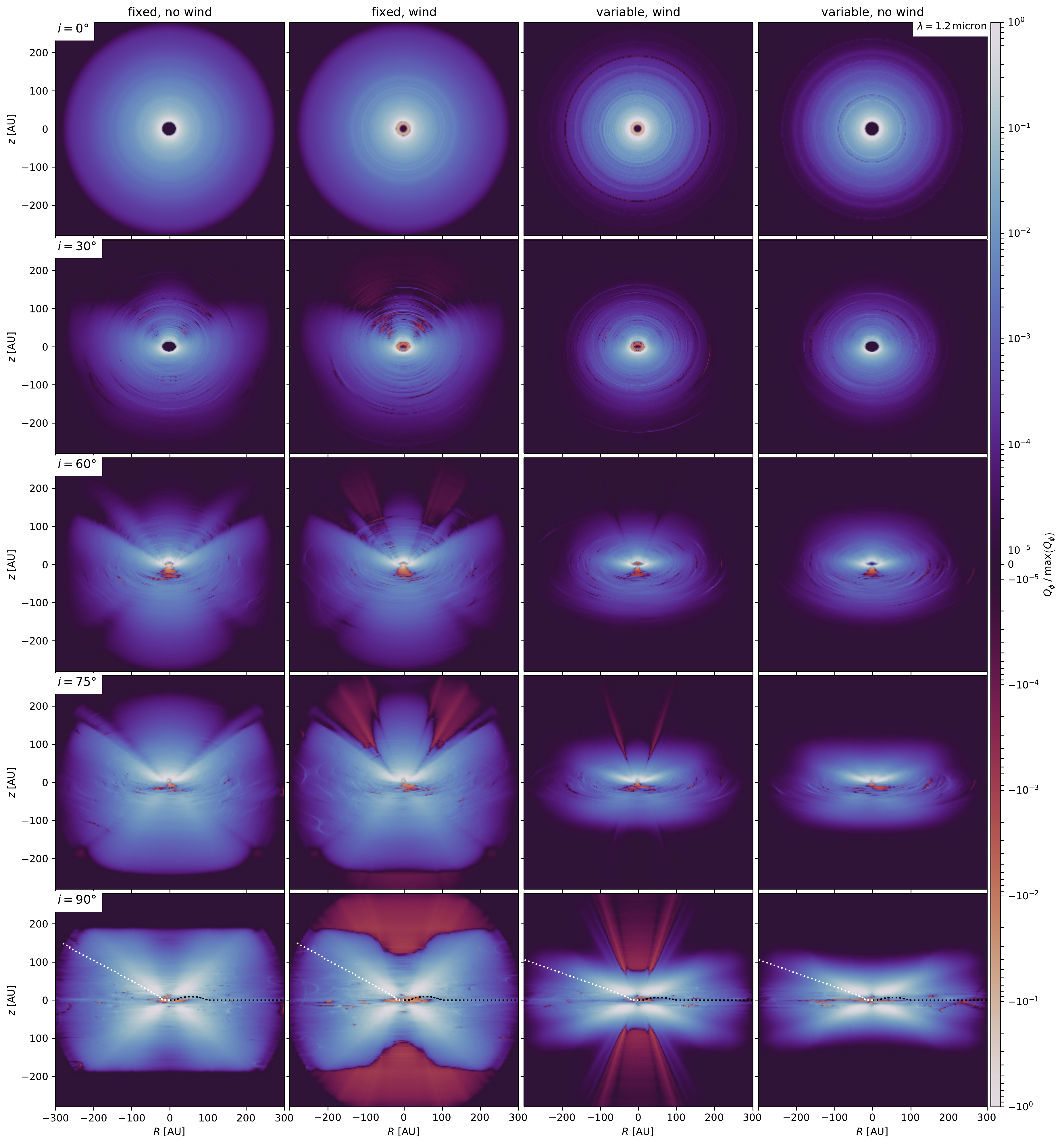}
    \caption{Polarised-light $Q_\phi$ for $\lambda_\mathrm{obs} = 1.2\,\mu$m for TD20.
    While the layout of the \textit{rows} (inclinations) and \textit{columns} (models) is the same as for the scattered-light images, no artificial coronagraph is applied.
    The dotted white (black) line indicates the $(\tau = 1)$-surface as seen from $r = 0$ ($z = \infty$).
    The wind produces a distinct signature especially around the jet region.}
    \label{fig:radmc-qphi-20-1.2}
\end{figure*}

\begin{figure*}
    \centering
    \includegraphics[width=0.99\textwidth]{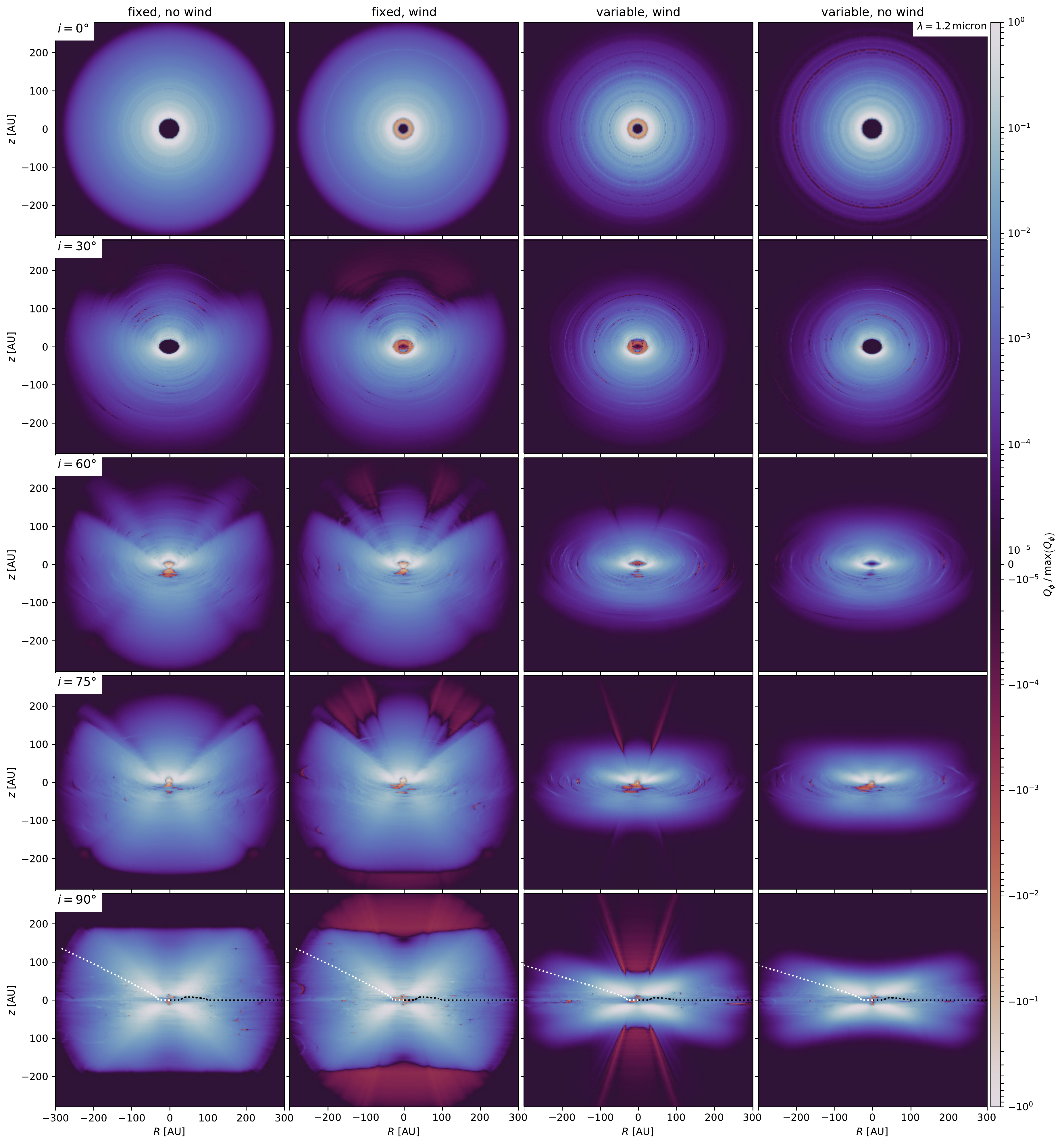}
    \caption{Polarised-light $Q_\phi$ for $\lambda_\mathrm{obs} = 1.2\,\mu$m for TD30, all else equal to Fig.~\ref{fig:radmc-qphi-20-1.2}, the characteristics of the wind signature included.}
    \label{fig:radmc-qphi-30-1.2}
\end{figure*}

The \texttt{RadMC-3D} results for $Q_\phi$ in polarised light at $\lambda_\mathrm{obs} = 1.2\,\mu$m are shown in Figs.~\ref{fig:radmc-qphi-20-1.2} (TD20) and \ref{fig:radmc-qphi-30-1.2} (TD30); no artificial coronagraph has been applied to these images.
Just like for the scattered-light intensities, we see that the inner hole of the transition disks appears much smaller in the presence of a photoevaporative wind.

Furthermore, the wind causes an area of $Q_\phi < 0$ above (and below) the bulk of the disk, which even cuts somewhat into the high-$z$, low-$R$ regions that otherwise exhibit $Q_\phi > 0$.
This feature is rather broad in the `fixed' models, and more spatially confined in the `variable' ones; it resembles the overall cone shape of the wind signature of the scattered-light images (see Figs.~\ref{fig:radmc-sci-20-0.7}--\ref{fig:radmc-sci-30-1.6}).
In the `fixed' setups, it can already be seen at $i = 30^\circ$; apart from the inner hole radius, this is the only outflow feature occurring at $i \lesssim 30^\circ$.

The maximum relative intensities of the wind signal (i.e. the cone feature) are $Q_\phi / \max(Q_\phi) \lesssim 10^{-2}$ for $\lambda_\mathrm{obs} \in \lbrace 0.4, 0.7, 1.2, 1.6 \rbrace \,\mu$m.\footnote{These values were retrieved for varTD20 at $i=90^\circ$, which gives the strongest cone feature. About one to two orders of magnitude need to be added if looking at different models and/or lower $i$.}
This stands in contrast to the strong fall-off of the relative brightness of the features with $\lambda_\mathrm{obs}$ seen in scattered light.

\subsubsection{Synthetic observations for SPHERE IRDIS}
\label{sec:Results:pol:irdis}

\begin{figure*}
    \centering
    \includegraphics[width=0.99\textwidth]{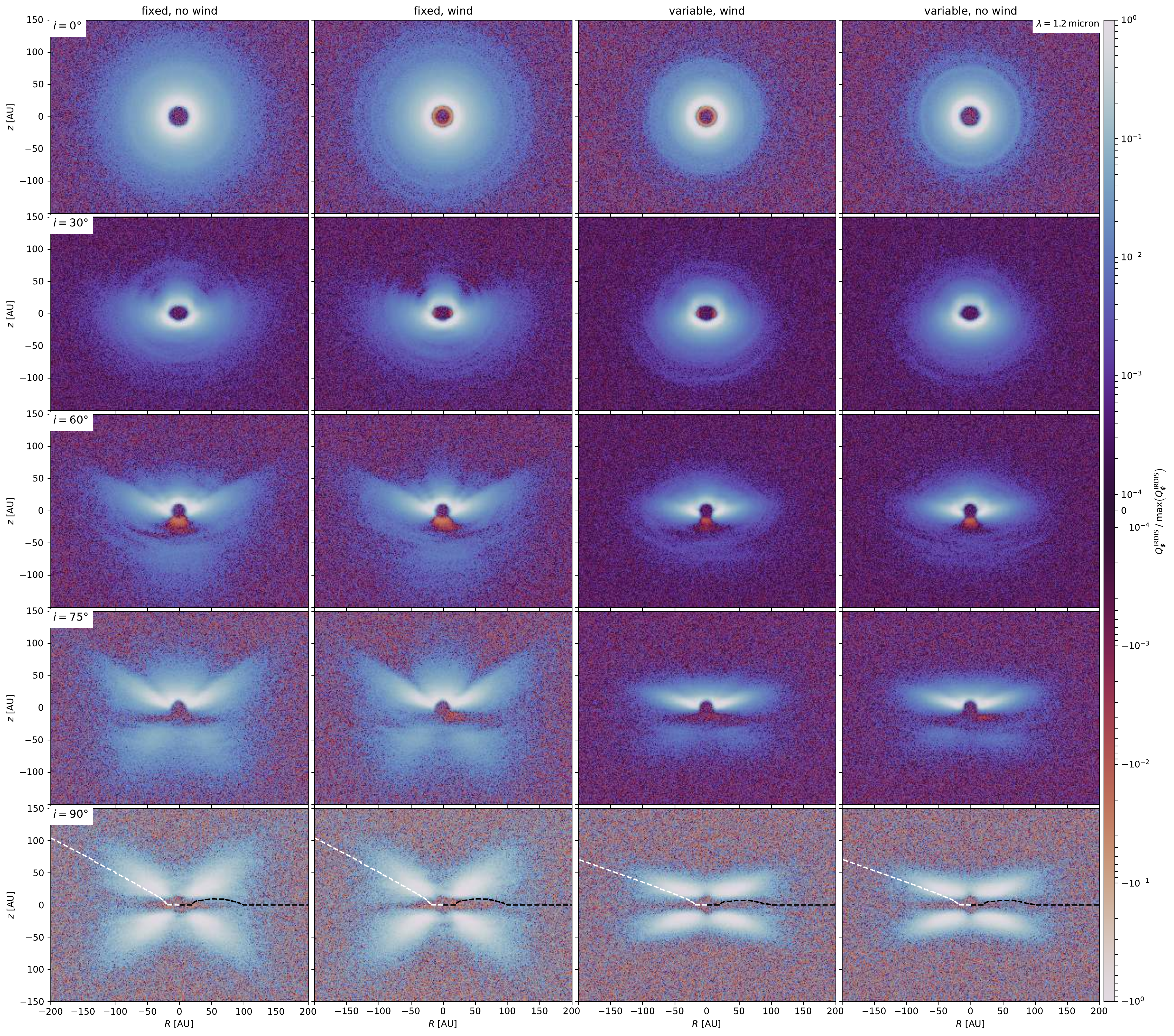}
    \caption{Synthesised $Q_\phi$ observations of TD20 in SPHERE IRDIS's $J$-band.
    The dotted white and black lines portray the same $(\tau = 1)$-surfaces as in Fig.~\ref{fig:radmc-qphi-20-1.2}.
    The instrument noise overshadows the wind features seen there.}
    \label{fig:irdis-20}
\end{figure*}

\begin{figure*}
    \centering
    \includegraphics[width=0.99\textwidth]{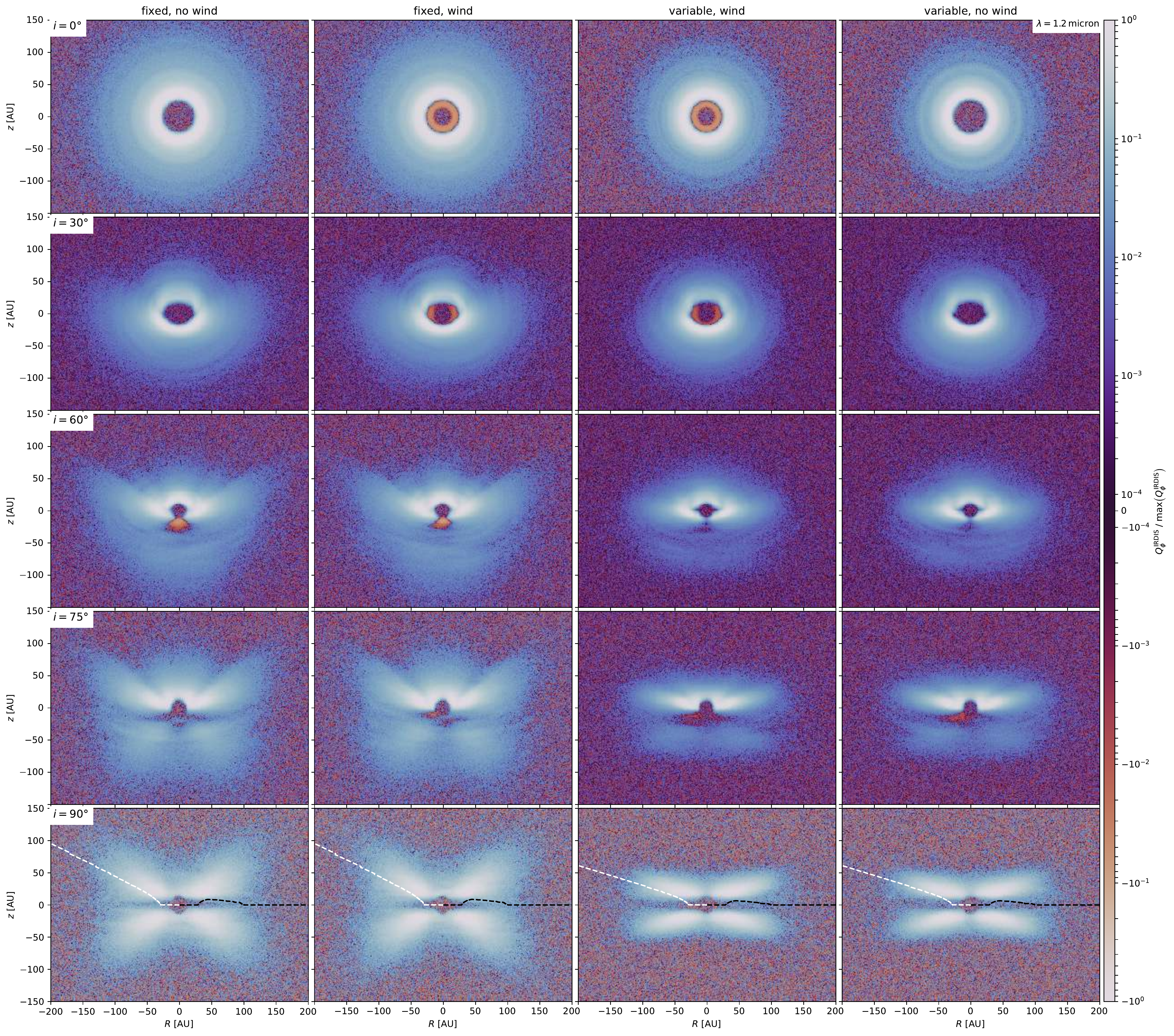}
    \caption{Synthesised $Q_\phi$ observations of TD30 in SPHERE IRDIS's $J$-band, all else equal to Fig.~\ref{fig:irdis-20}.}
    \label{fig:irdis-30}
\end{figure*}

The synthesised instrument responses for SPHERE IRDIS's $J$-band are shown in Figs.~\ref{fig:irdis-20} (TD20) and \ref{fig:irdis-30} (TD30).
The wind features seen in Figs.~\ref{fig:radmc-qphi-20-1.2} and \ref{fig:radmc-qphi-30-1.2} are drained in instrument noise; this is also the case for $H$-band, so we do not include the corresponding plots.

While the coronagraph of SPHERE IRDIS obscures most of the difference in inner hole sizes found in Sect.~\ref{sec:Results:pol:radmc}, an inner ring of $Q_\phi < 0$ remains at $i \lesssim 30^\circ$; probably due to the larger gap size, it is more prominent for TD30 than TD20.
The cone-like features seen in the clear $Q_\phi$ images (Figs.~\ref{fig:radmc-qphi-20-1.2} and \ref{fig:radmc-qphi-30-1.2}) are much less pronounced in the synthetic observations.
They arguably still appear at intermediate inclinations (especially $30^\circ \lesssim i \lesssim 60^\circ$), and are more pronounced for the `fixed' models.
Further post-processing with a dedicated noise removal tool such as \texttt{denoise} \citep{Price-2007} did not enhance the wind signatures.\footnote{\texttt{denoise}: \href{https://github.com/danieljprice/denoise/}{[link]}}

\begin{figure*}
    \centering
    \includegraphics[width=0.6\textwidth]{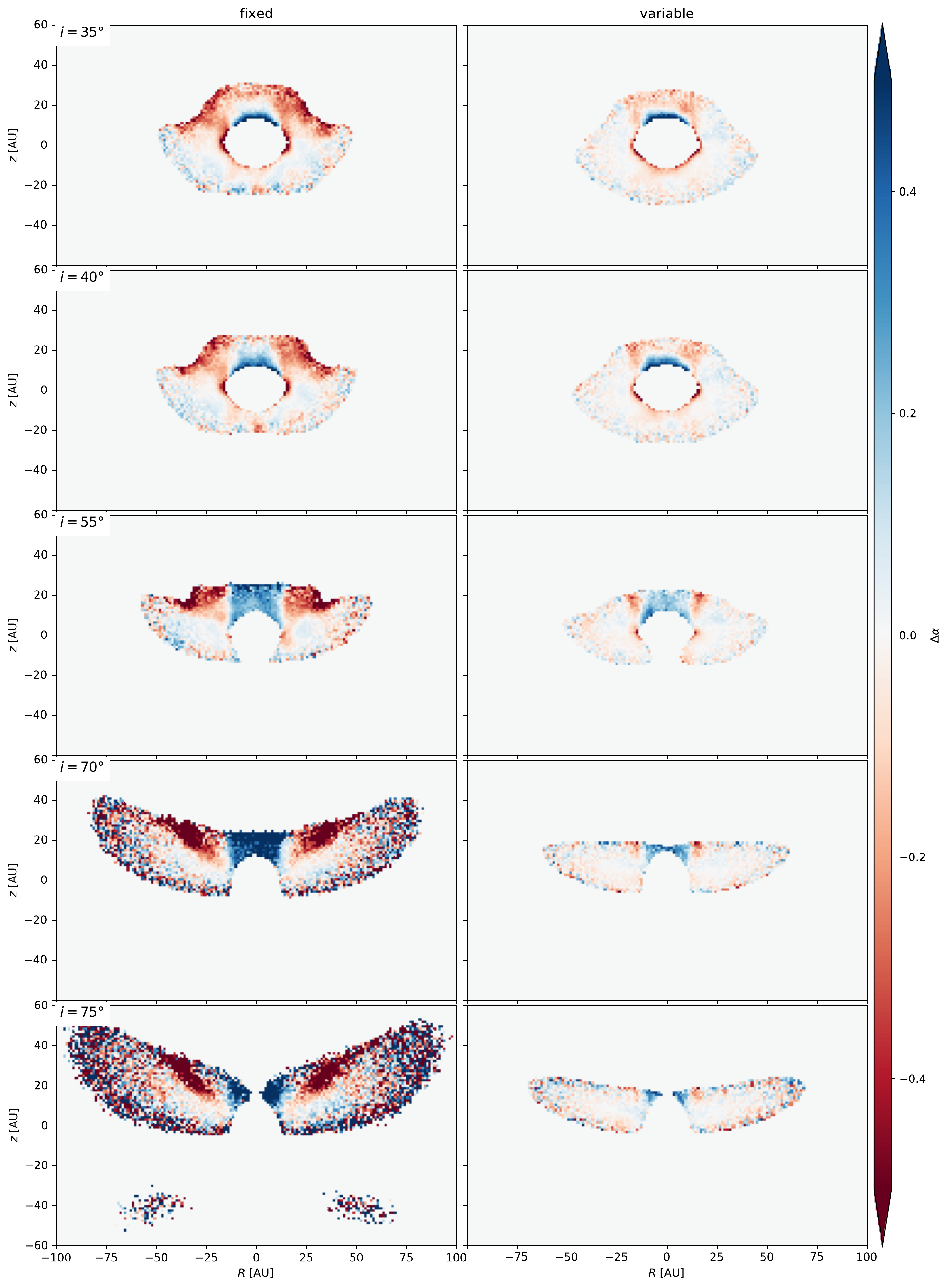}
    \caption{Difference $\Delta \alpha$ between the spectral indices $\alpha_{J,H}$ of the `wind' and `no wind' models for fixTD20 (\textit{left}) and varTD20 (\textit{right}); plots for the individual $\alpha$ values are provided in App.~\ref{sec:app:images}.
    For intermediate inclinations ($35^\circ \lesssim i \lesssim 75^\circ$), the dusty wind causes a clear colour excess, which appears particularly blue above the star.}
    \label{fig:irdis-20-DeltaAlpha}
\end{figure*}

By contrast, the difference in spectral indices $\Delta \alpha$ (see Sect.~\ref{sec:Methods:images:pol}) extracted from the SPHERE IRDIS $P$ predictions is clearly non-zero.
In the plots of Fig.~\ref{fig:irdis-20-DeltaAlpha}, we see a distinct blue excess above the location of the star.
It occurs for $35^\circ \lesssim i \lesssim 75^\circ$; for clarity and to avoid high noise levels, regions with a weak predicted signal ($P < 0.04 \cdot \max(P)$) were masked out before computing $\alpha$.
This indicates that above the star, $\alpha_{J,H}$ is greater in the `wind' than in the `no wind' cases, and thus that an XEUV wind enhances the difference in $P$ between $J$- and $H$-bands.

%%%%%%%%%%%%%%%%%%%%%%%%%%%%%%%%%%%%%%%%%%%%%%%%%%%%%%%%%%%%%%%%%%%%%%%%%%%%%%%%%%%%%%%%%%%%%%%%%%%%

\section{Discussion}
\label{sec:Discussion}

\subsection{Scattered light}
\label{sec:Discussion:sca}

The vertical settling of the dust has a significant impact on the observability of a dusty outflow; if the disk is assumed to be large and the settling to be minimal (the `fixed' case), large dust densities at high $z$ are required for a noticeable signal.
For a smaller disk or stronger vertical settling, less dust is needed for a current instrument to be able to pick up the signature; this is due to the radial trajectories on which even the larger entrainable grains eventually leave the stellar gravity well, and which mean that the dust densities decline with $r$.

In Paper~\citetalias{Franz-2020}, we suggested that $\left. \max(z) \right|_R$ may be usable as a tracer for an XEUV wind, and thus allow one to distinguish it from a scenario dominated by an MHD wind.
This did not work in Paper~\citetalias{Franz-2022a}, and also does not work here.
Comparing the radiative-transfer results for $\lambda_\mathrm{obs} = 0.7\,\mu$m and $1.6\,\mu$m (see Figs.~\ref{fig:radmc-sci-20-0.7}--\ref{fig:radmc-sci-30-1.6}) does not show a big difference between the opening angles of the wind-induced cone features; this is despite the `variable' models having a strong concentration of entrained dust towards $\left. \max(z) \right|_R$ (see Figs.~\ref{fig:dust-rho-20-fix}--\ref{fig:dust-rho-30-var}) for almost all $a_0$.

While we are not aware of any clear observations to date which show an outflow signature as suggested by these \texttt{RadMC-3D} results, the synthesised coronagraphic images for JWST NIRCam imply that imaging dusty winds will soon be a realistic possibility.
Even if we have set up our models in Paper~\citetalias{Franz-2022a} and here to present best-case scenarios, the wind signatures seen in the synthesised observations (Figs.~\ref{fig:jwst-20} and \ref{fig:jwst-30}) are prominent especially for the `variable' model.
In addition, while trying to maximise the signal, we kept the dust-to-gas ratio at 0.01; while this value is generally used \citep[see e.g.][]{Andrews-2009, Andrews-2010}, it may underestimate the actual dust content of the disk as noted in Sect.~\ref{sec:Methods:dust:rho}.

For more face-on transition disks, the intensity profile of the inner hole may allow a tentative distinction between disks with and without dusty winds, as long as an estimate of the gap size can be retrieved by other means and there are no other dusty outflows (e.g. magnetically-driven jets).
Planetesimals and planets (which are likely present once the disk enters its transition stage) may also strongly impact the radial brightness profile; the differences between planet-carved and photoevaporative gap intensity profiles will be investigated in a future work (Sch{\"a}fer et al., in prep.).

\subsection{Polarised light}
\label{sec:Discussion:pol}

While the clean $Q_\phi$ images (Figs.~\ref{fig:radmc-qphi-20-1.2} and \ref{fig:radmc-qphi-30-1.2}) show a clear wind signature, the noise in the synthesised observations for SPHERE IRDIS (Figs.~\ref{fig:irdis-20} and \ref{fig:irdis-30}) mostly obfuscates it.
As is shown in App.~\ref{sec:app:images}, a reduction in instrument noise by a factor $\lesssim 10$ would allow for the detection of an observational signature, in particular at intermediate $i$.
The narrow spatial extent of the high-intensity $(Q_\phi < 0)$-regions seen in Figs.~\ref{fig:radmc-qphi-20-1.2} and \ref{fig:radmc-qphi-30-1.2} (especially for the `variable' model) may render their detection unfeasible, considering that the contrast achievable with SPHERE IRDIS drops off at small angles \citep[see e.g.][their Fig.~5]{Boccaletti-2008}.
In that case, it may be worth looking into which coronagraphic pupil delivers the best contrast at small angles.\footnote{Due to the many other uncertainties of the model, we decided to not explore this avenue further in this work.}

Furthermore, analyses of $Q_\phi$ data are often performed in linear stretch in the literature.
Due to the low relative intensities of the material in the wind, switching to logarithmic stretch may reveal interesting additional features \citep[see e.g.][and Paper~II]{Avenhaus-2018}.
In the case of very strong dust entrainment in specific systems, signatures like the ones showcased here may be present.
Speculatively, this could also be the case for MY Lup (see Paper~\citetalias{Franz-2022a}, Fig.~11); but whereas the analyses of \citet[][using X-shooter on the VLT]{Alcala-2017} suggest low accretion rates and thus a cleared inner hole similar to the models shown here (but maybe with a smaller $r_\mathrm{gap}$), later data retrieved by \citet[][using HST]{Alcala-2019} seem more in line with an intact inner disk.

The polarised signal may be more prominent when observing at smaller wavelengths.
The Zurich Imaging Polarimeter (ZIMPOL) of SPHERE would be able to do exactly that, however we do not yet have a noise profile for this instrument.
Furthermore, analyses as performed for instance by \citet{Thalmann-2015} and \citet{deBoer-2017} do not find clear outflow signatures.
This may in part be due to most current investigations using data from SPHERE ZIMPOL to investigate systems at lower inclinations with regards to planet(esimal)s or companions \citep[e.g.][]{deBoer-2016, Stolker-2016, Avenhaus-2017, Bertrang-2018, Cugno-2019, Willson-2019}.

Once a set of disks with and without confirmed dusty winds has been created from SPHERE ZIMPOL observations and (deep) SPHERE IRDIS data, the retrieved colours could then be employed to establish a baseline for the expected signal strength and morphology in polarised light.
The colour excesses shown in Fig.~\ref{fig:irdis-20-DeltaAlpha} could then be used to determine whether a wind is present in other sources.

\subsection{Caveats and outlook}
\label{sec:Discussion:limits}

As noted above, this work aims to present a numerically simplified, best-case scenario for the observability of XEUV winds, in order to investigate whether they could be detected via $\mu$m dust observations with modern instruments at all.
The `fixed' model represents the simplest setup possible, and is illustrative as it allows to decouple the effect of the wind from the complex dust evolution processes that happen in the disk.
Studying the differences between the `fixed' and `variable' models provides insights as to what wind features may become more prominent when different physical processes drive the dust distribution at the launch region; an example is the clear enhancement of the narrow outflow channels in the case of dust settling.
Nonetheless, our results may well overestimate the dust content of the photoevaporation-driven outflow.

The presence of a gas pressure bump at the outer gap edge may invalidate the assumption of a direct relation between $\varrho_\mathrm{gas}$ and $\varrho_\mathrm{dust}$, which could reduce the dust densities in the wind, but could also enhance the local dust-to-gas ratio \citep[see e.g.][]{Garate-2021}.
Further studies with a more realistic treatment of the outer gap edge will be needed to more accurately assess this aspect.

In addition, the dust evolution in the disk has been assumed to be fully decoupled from the wind model.
The fast evolution of $\dot{M}_\mathrm{dust} / \dot{M}_\mathrm{gas}$, which would exceed unity in $\approx 0.1\,$Myr assuming a steady-state case, is a direct consequence of this, and as such should be treated with care; thus, dust densities in the wind may be lower than predicted here, or may vary over time.
Both of these options would reduce the observability of the features presented.

The outflowing dust must be replenished at the disk surface; for the low-$z$ environment of the outer gap edge, which accounts for the main portion of $\dot{M}_\mathrm{dust}$, the dominant process for this is radial drift.
A quick estimate of the required drift velocities via $v_r = \dot{M}_\mathrm{dust} / (2\,\pi\, R \, \Sigma_\mathrm{dust})$ yields $v_r \approx 10...0.1\,$m/s ($2 \cdot 10^{-3}...2 \cdot 10^{-5}\,$AU/yr) for $a_0 = 0.01...12\,\mu$m; here, $R$ has been chosen at $\max(\Sigma_\mathrm{gas})$, that is $R \simeq 26\,$AU for TD20 and $R \simeq 40\,$AU for TD30.
\citet{Kanagawa-2017} have reported drift speeds of up to $5.5 \cdot 10^{-3}\,$AU/yr for $\alpha=10^{-3}$ and a dust-to-gas ratio of 0.01; but these values were found for $St \approx 1$, whereas $\mu$m-sized grains have $St \ll 1$ in the midplane.
Another question would be the amount of dust already present at the gap edge, which may function as a buffer if material inflow from farther out is rather low.
So in conclusion, it is questionable whether the $\dot{M}_\mathrm{dust}$ of Table~\ref{tab:massloss} can be sustained over prolonged periods of time, or the outflow is for instance somewhat periodic; this will need to be investigated in a follow-up study.

%%%%%%%%%%%%%%%%%%%%%%%%%%%%%%%%%%%%%%%%%%%%%%%%%%%%%%%%%%%%%%%%%%%%%%%%%%%%%%%%%%%%%%%%%%%%%%%%%%%%

\section{Summary}
\label{sec:Summary}

In this work, we have modelled the XEUV-driven dusty outflow for two transition disks with inner hole radii of about 20 and 30\,AU, and produced synthetic observations to predict their observability in scattered light with JWST NIRCam and in polarised light with SPHERE IRDIS.
Throughout the modelling process, we made assumptions as to provide a best-case scenario for the visibility of the wind.
Our findings can be summarised as follows:

\begin{itemize}
    \item For a uniform dust-to-gas ratio, the dust mass outflow is still uniform as for a primordial disk.
    If dust settling is accounted for, preferred outflow channels emerge especially for the largest entrainable grains.
    
    \item These preferred outflow channels produce a distinct wind signature, consisting of a cone-shaped feature around the polar axis of the disk, in both scattered and polarised light.
    The feature is brighter at higher inclinations.
    
    \item JWST NIRCam should be able to detect a dusty photoevaporative outflow if the underlying disk is similar to the models presented here (i.e. $M_* \approx 0.7\,\mathrm{M}_\odot$, $L_X \approx 2 \cdot 10^{30}\,\mathrm{erg/s}$, $M_\mathrm{disk} \approx 5 \cdot 10^{-3} \, M_*$, dust-to-gas ratio of 0.01, gap size of 20...30\,AU; see Sect.~\ref{sec:Methods:dust}) and the observational set-up is appropriately chosen.
    This also applies if no wind signature has been detected with SPHERE IRDIS.
    
    \item Compared to wind-less disks, dusty winds cause a colour excess in polarised light.
\end{itemize}

For more realistic modelling, further studies are required to better constrain the dust densities at the outer gap edge of transition disks; depending on the dust reservoir, steady or intermittent dusty wind signatures may be possible.
In addition, deep observations with current instruments can be employed to try and identify objects with an XEUV-driven dusty outflow.

%%%%%%%%%%%%%%%%%%%%%%%%%%%%%%%%%%%%%%%%%%%%%%%%%%%%%%%%%%%%%%%%%%%%%%%%%%%%%%%%%%%%%%%%%%%%%%%%%%%%

\begin{acknowledgements}

We would like to thank P.~Rodenkirch, P.-G.~Valeg{\aa}rd, M.~G{\'a}rate, and P.~Weber for helpful discussions, and the (anonymous) referee for a constructive report that improved the manuscript.
\\
Furthermore, we thank C.~Dullemond for providing a pre-release version of the \texttt{disklab} package together with T.B.
\\
This research was funded by the Deutsche Forschungsgemeinschaft (DFG, German Research Foundation), grant 325594231 (FOR 2634/1 and FOR 2634/2), and the Munich Institute for Astro- and Particle Physics (MIAPP) of the DFG cluster of excellence \textit{Origin and Structure of the Universe}.
B.E.~and T.B.~acknowledge funding by the DFG under Germany's Excellence Strategy -- EXC-2094-390783311.
S.C.~and S.P.~acknowledge support from Agencia Nacional de Investigaci{\'o}n y Desarrollo de Chile (ANID) through FONDECYT Regular grants 1211496 and 1191934.
T.B.~acknowledges funding from the European Research Council (ERC) under the European Union's Horizon 2020 research and innovation programme under grant agreement No 714769, as well as funding by the DFG under grant 361140270.
CHR is grateful for support from the Max Planck Society. 
The simulations have mostly been carried out on the computing facilities of the Computational Center for Particle- and Astrophysics (C2PAP).

\end{acknowledgements}

%%%%%%%%%%%%%%%%%%%%%%%%%%%%%%%%%%%%%%%%%%%%%%%%%%%%%%%%%%%%%%%%%%%%%%%%%%%%%%%%%%%%%%%%%%%%%%%%%%%%

\bibliographystyle{aa}
\bibliography{Literature.bib}

%%%%%%%%%%%%%%%%%%%%%%%%%%%%%%%%%%%%%%%%%%%%%%%%%%%%%%%%%%%%%%%%%%%%%%%%%%%%%%%%%%%%%%%%%%%%%%%%%%%%

\begin{appendix}

{\onecolumn

\section{The impact of the opacity}
\label{sec:app:opacities}

For this work as well as Paper~\citetalias{Franz-2022a}, we used an opacity prescription assuming DSHARP values for the disk and pure astrosilicates for the wind (see Sect.~\ref{sec:Methods:images}).
To test the impact of the opacity prescription, we performed additional radiative transfer simulations for the `wind' models of TD20 at $\lambda_\mathrm{obs} = 1.2\,\mu$m and $i = 90^\circ$, using different opacities.
In these simulations, both the disk and wind regions of the model employ the same opacity prescription.
The opacities tested were, firstly, pure astrosilicates \citep{Draine-2003c}, secondly, DSHARP values \citep[][and references therein]{Birnstiel-2018}, and thirdly, Ricci opacities \citep[][and references therein]{Ricci-2010a}.\footnote{The Ricci opacities are computed for $\varrho_\mathrm{grain} = 1.2\,\mathrm{g/cm^3}$; while this differs from our usual value of $\varrho_\mathrm{grain} = 1\,\mathrm{g/cm^3}$, the difference should be small enough to not significantly impact the overall results.}
The results are shown in Figs.~\ref{fig:opac-sci-fix} (fixTD20) and \ref{fig:opac-sci-var} (varTD20).

\begin{figure*}[!h]
    \centering
    \includegraphics[width=0.99\textwidth]{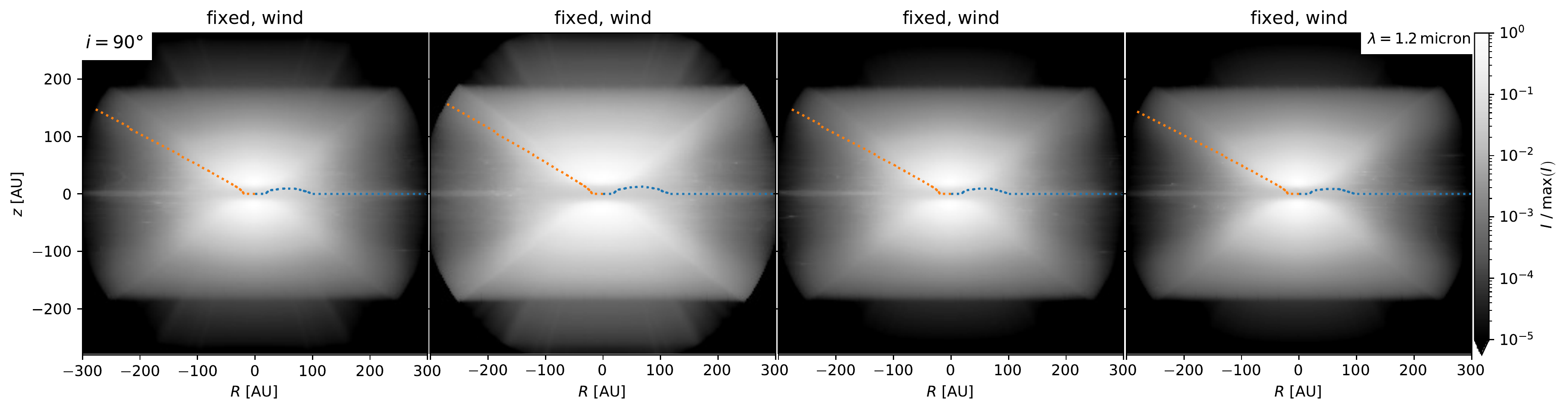}
    \caption{Scattered-light intensities for fixTD20, $\lambda_\mathrm{obs} = 1.2\,\mu$m, $i = 90^\circ$.
    The panels show, from left to right, the results for our opacity mix, pure astrosilicates, DSHARP opacities \citep{Birnstiel-2018}, and \citet{Ricci-2010a} opacities.
    The coloured lines indicate the ($\tau = 1$)-surfaces as in Fig.~\ref{fig:radmc-sci-20-0.7}.}
    \label{fig:opac-sci-fix}
\end{figure*}

\begin{figure*}[!h]
    \centering
    \includegraphics[width=0.99\textwidth]{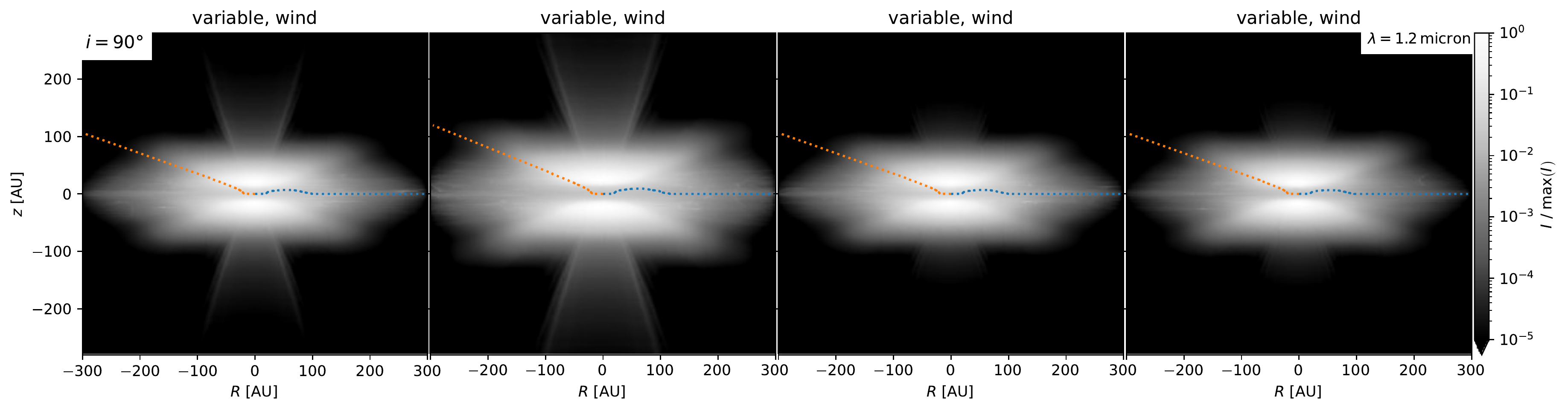}
    \caption{Scattered-light intensities for varTD20, $\lambda_\mathrm{obs} = 1.2\,\mu$m, $i = 90^\circ$; all else equal to Fig.~\ref{fig:opac-sci-fix}.}
    \label{fig:opac-sci-var}
\end{figure*}

Expectedly, assuming a silicate-only material composition yields the highest intensities for both the disk and wind regions; switching to DSHARP opacities for only the disk (as done in the main part of this work) slightly reduces the signal from the dusty outflow.
Conversely, using DSHARP-only or Ricci-only opacities strongly reduces the signal caused by the XEUV wind; the latter results appear quite similar, despite the difference in absorption opacities between the prescriptions.\footnote{At least for smooth disks, Ricci opacities might be more realistic than DSHARP ones, see \citet{Zormpas-2022}.}
All in all, the opacity prescription used in this work (and Paper~\citetalias{Franz-2022a}) fits its purpose of providing a best-case scenario.

In addition, comparing (highly detailed) observational data in polarised light to models with various opacities may help to identify the most realistic model for the dust composition and settling.
This can be seen from the varying signs of $Q_\phi$ in Figs.~\ref{fig:opac-qphi-fix} and \ref{fig:opac-qphi-var}, where the only parameter varied between the individual subplots is their opacity model.

\begin{figure*}[!h]
    \centering
    \includegraphics[width=0.99\textwidth]{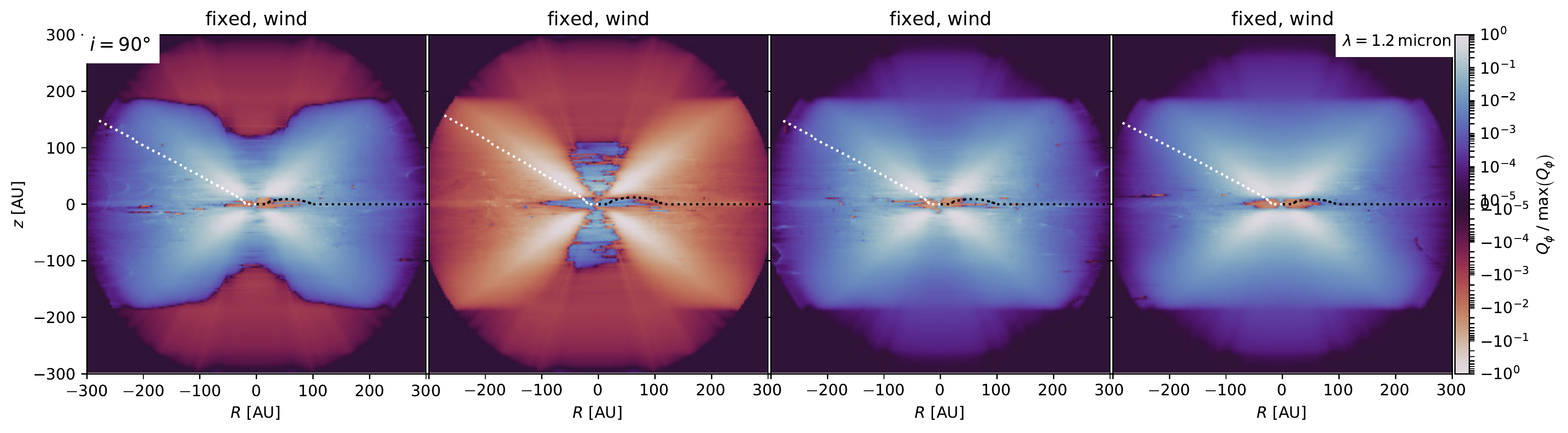}
    \caption{Polarised-light $Q_\phi$ for fixTD20, $\lambda_\mathrm{obs} = 1.2\,\mu$m, $i = 90^\circ$.
    The panels show, from left to right, the results for our opacity mix, pure astrosilicates, DSHARP opacities \citep{Birnstiel-2018}, and \citet{Ricci-2010a} opacities.
    The coloured lines indicate the ($\tau = 1$)-surfaces as in Fig.~\ref{fig:radmc-qphi-20-0.7}.}
    \label{fig:opac-qphi-fix}
\end{figure*}

\begin{figure*}[!h]
    \centering
    \includegraphics[width=0.99\textwidth]{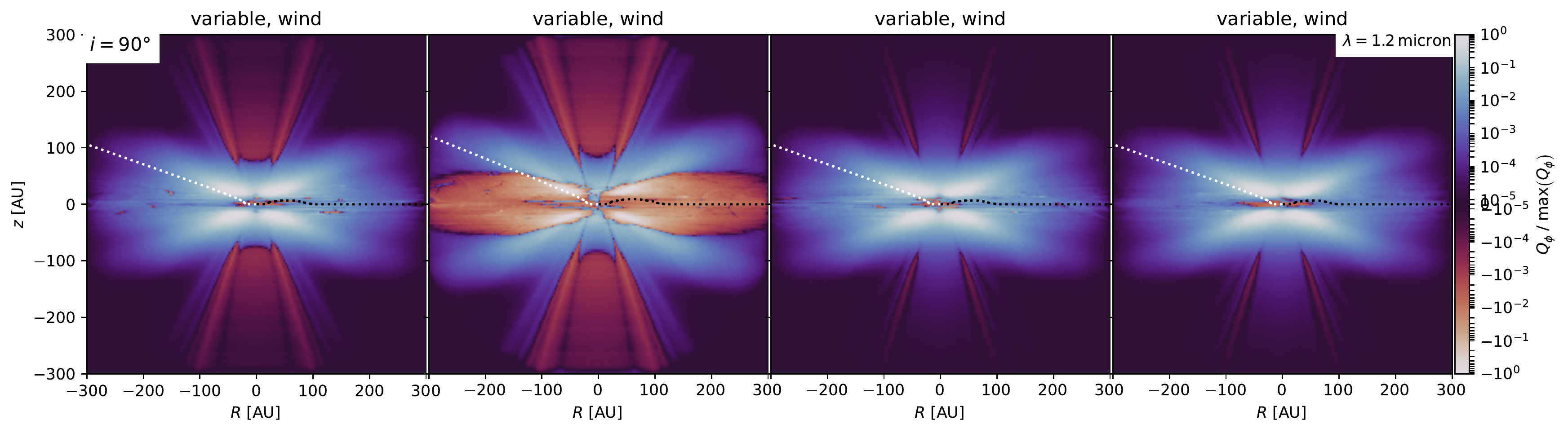}
    \caption{Polarised-light $Q_\phi$ for varTD20, $\lambda_\mathrm{obs} = 1.2\,\mu$m, $i = 90^\circ$; all else equal to Fig.~\ref{fig:opac-qphi-fix}.}
    \label{fig:opac-qphi-var}
\end{figure*}

\newpage

%%%%%%%%%%%%%%%%%%%%%%%%%%%%%%%%%%%%%%%%%%%%%%%%%%%%%%%%%%%%%%%%%%%%%%%%%%%%%%%%%%%%%%%%%%%%%%%%%%%%

\section{Additional images}
\label{sec:app:images}

For clarity, we have not included all our radiative-transfer results in Sect.~\ref{sec:Results}; however, some of them may still be of interest for future observational campaigns.
Figs.~\ref{fig:radmc-sci-20-1.2} and \ref{fig:radmc-sci-30-1.2} show the scattered-light intensities for $\lambda_\mathrm{obs} = 1.2\,\mu$m.
As can be seen from a comparison between these images and Figs.~\ref{fig:radmc-sci-20-0.7}--\ref{fig:radmc-sci-30-1.6}, smaller $\lambda_\mathrm{obs}$ do not necessarily lead to stronger wind features; this is the case especially for the `variable' model, probably due to the disk being more vertically settled for larger $a_0$.
At $\lambda = 1.2\,\mu$m and $i = 30^\circ$, varTD20 already exhibits a wind-driven cone feature.

\begin{figure*}
    \centering
    \includegraphics[width=0.99\textwidth]{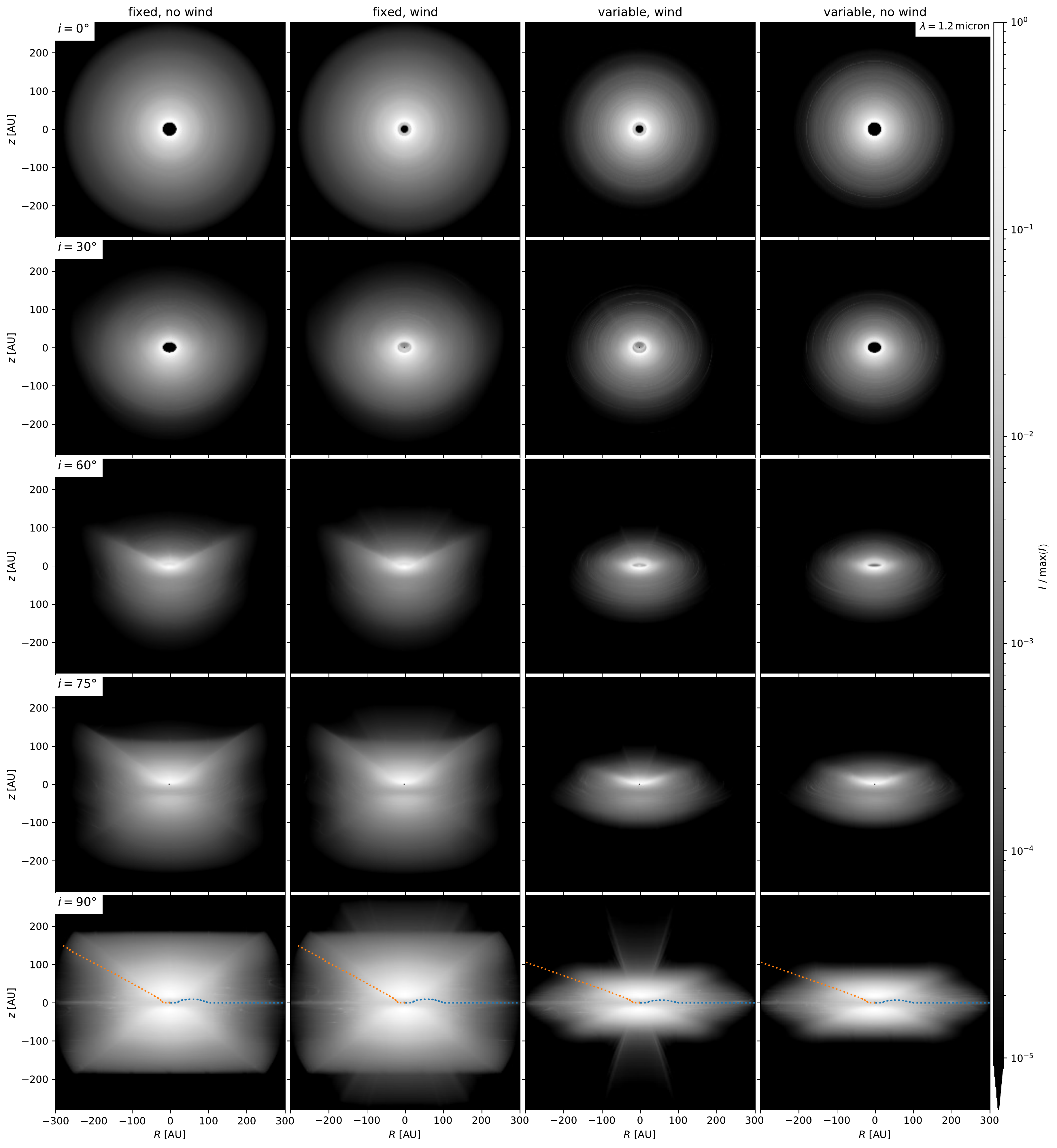}
    \caption{Scattered-light intensities for $\lambda_\mathrm{obs} = 1.2\,\mu$m for TD20, all else equal to Fig.~\ref{fig:radmc-sci-20-0.7}.}
    \label{fig:radmc-sci-20-1.2}
\end{figure*}

\begin{figure*}
    \centering
    \includegraphics[width=0.99\textwidth]{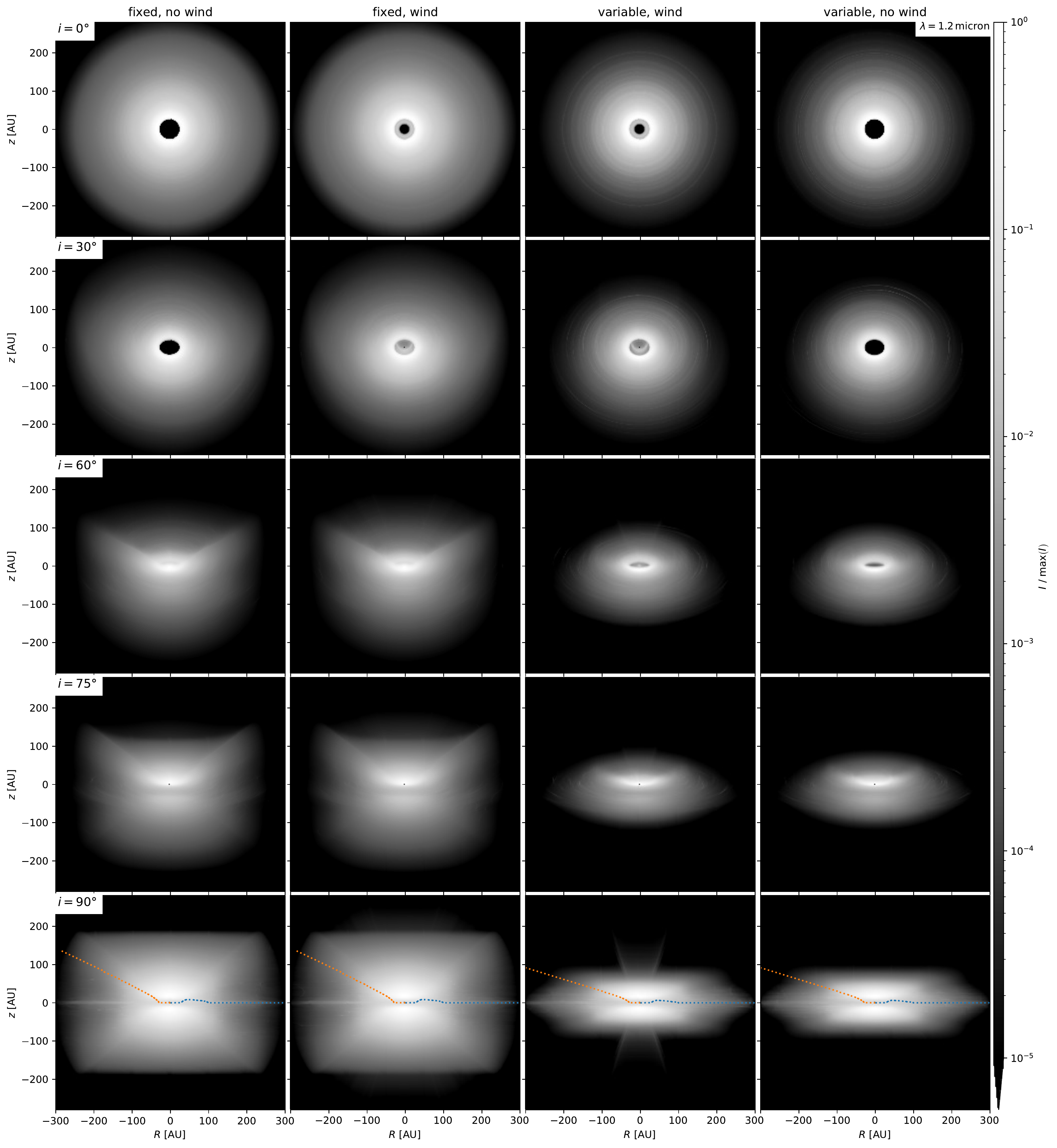}
    \caption{Scattered-light intensities for $\lambda_\mathrm{obs} = 1.2\,\mu$m for TD30, all else equal to Fig.~\ref{fig:radmc-sci-20-0.7}.}
    \label{fig:radmc-sci-30-1.2}
\end{figure*}

The $Q_\phi$ images for $\lambda_\mathrm{obs} = 0.7\,\mu$m, which may be interesting also for SPHERE ZIMPOL, can be seen in Figs.~\ref{fig:radmc-qphi-20-0.7} (TD20) and \ref{fig:radmc-qphi-30-0.7} (TD30); their counterparts for $\lambda_\mathrm{obs} = 1.6\,\mu$m are shown in Figs.~\ref{fig:radmc-qphi-20-1.6} (TD20) and \ref{fig:radmc-qphi-30-1.6} (TD30).
When including $\lambda_\mathrm{obs} = 1.2\,\mu$m (Figs.~\ref{fig:radmc-qphi-20-1.2} and \ref{fig:radmc-qphi-30-1.2}) for comparison, we see that the sign of the $Q_\phi$ signal of the XEUV-driven outflow changes between the plots for $0.7\,\mu$m and $1.2\,\mu$m.
In all cases, we do however retain the cone-shaped feature above the disk midplane.

\begin{figure*}
    \centering
    \includegraphics[width=0.99\textwidth]{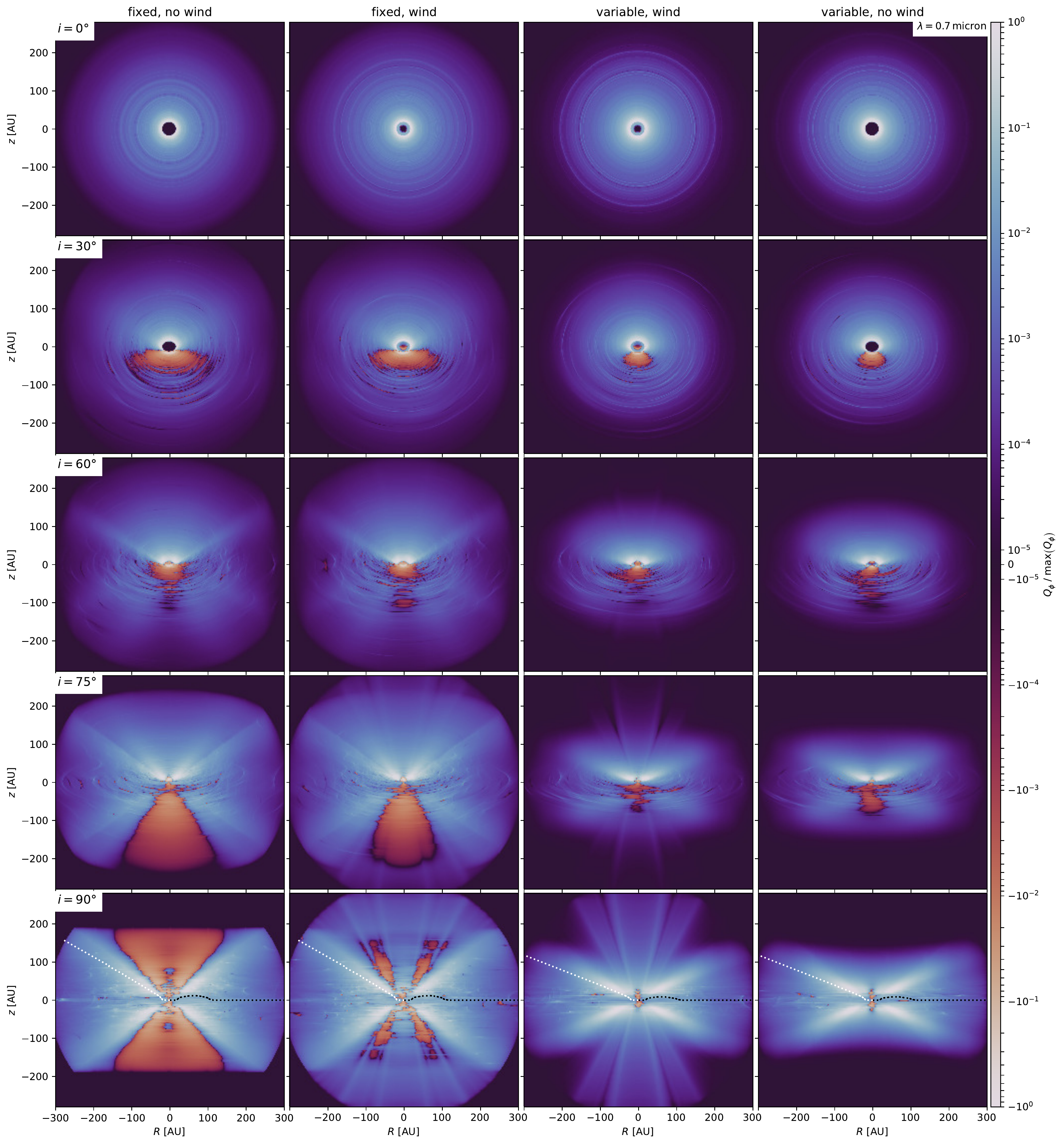}
    \caption{Polarised-light $Q_\phi$ for $\lambda_\mathrm{obs} = 0.7\,\mu$m for TD20, all else equal to Fig.~\ref{fig:radmc-qphi-20-1.2}.}
    \label{fig:radmc-qphi-20-0.7}
\end{figure*}

\begin{figure*}
    \centering
    \includegraphics[width=0.99\textwidth]{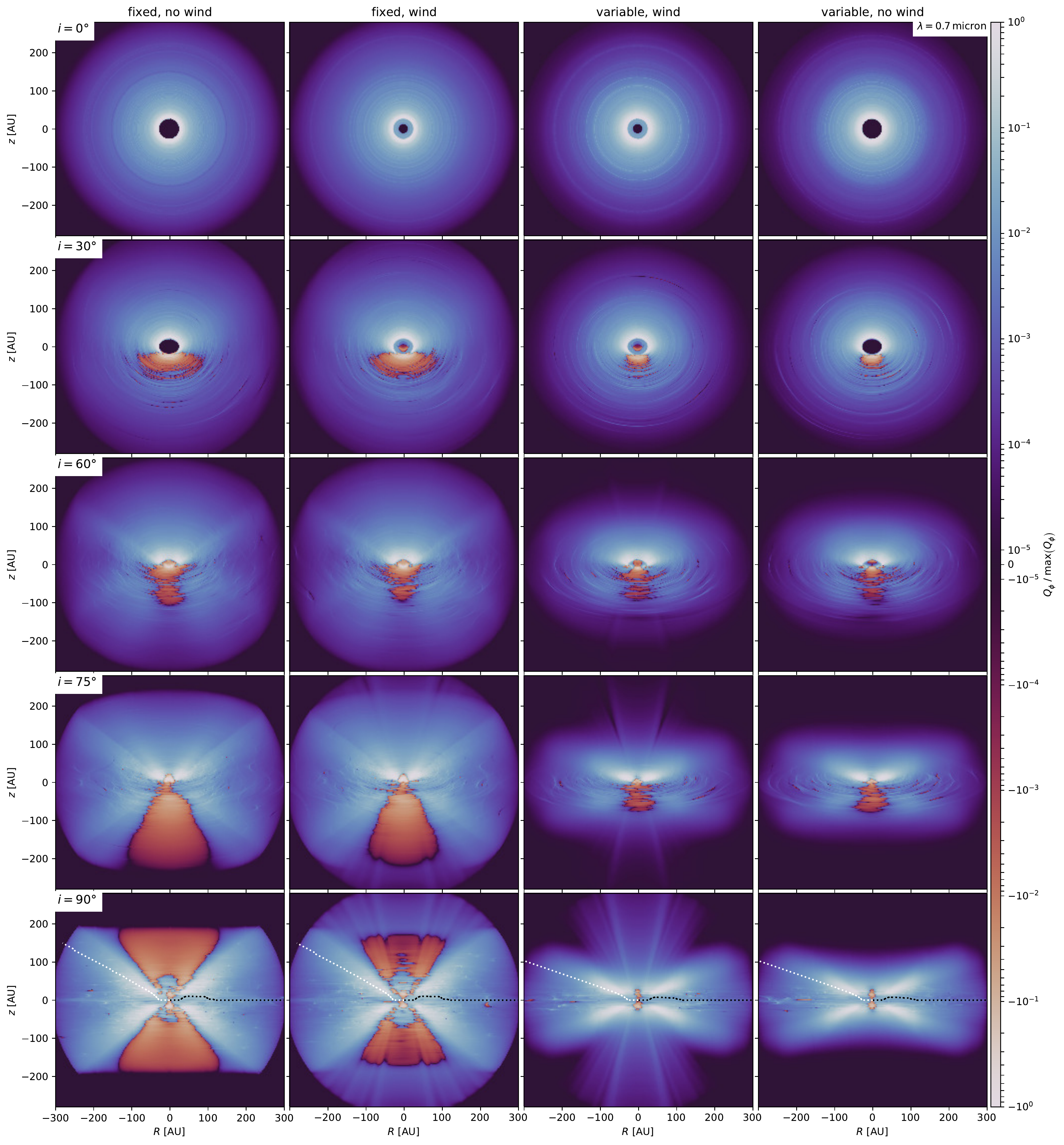}
    \caption{Polarised-light $Q_\phi$ for $\lambda_\mathrm{obs} = 0.7\,\mu$m for TD30, all else equal to Fig.~\ref{fig:radmc-qphi-20-1.2}.}
    \label{fig:radmc-qphi-30-0.7}
\end{figure*}

\begin{figure*}
    \centering
    \includegraphics[width=0.99\textwidth]{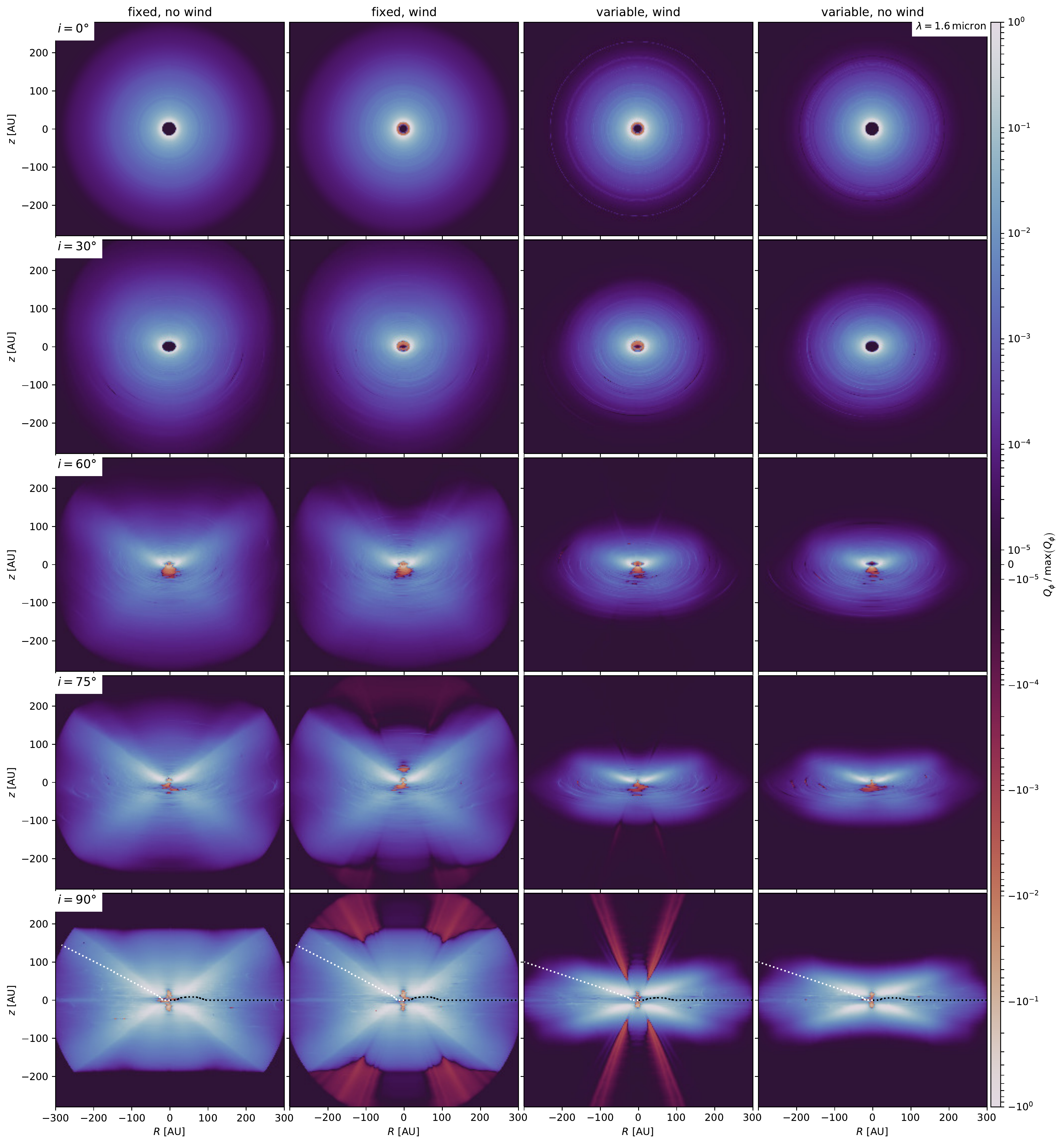}
    \caption{Polarised-light $Q_\phi$ for $\lambda_\mathrm{obs} = 1.6\,\mu$m for TD20, all else equal to Fig.~\ref{fig:radmc-qphi-20-1.2}.}
    \label{fig:radmc-qphi-20-1.6}
\end{figure*}

\begin{figure*}
    \centering
    \includegraphics[width=0.99\textwidth]{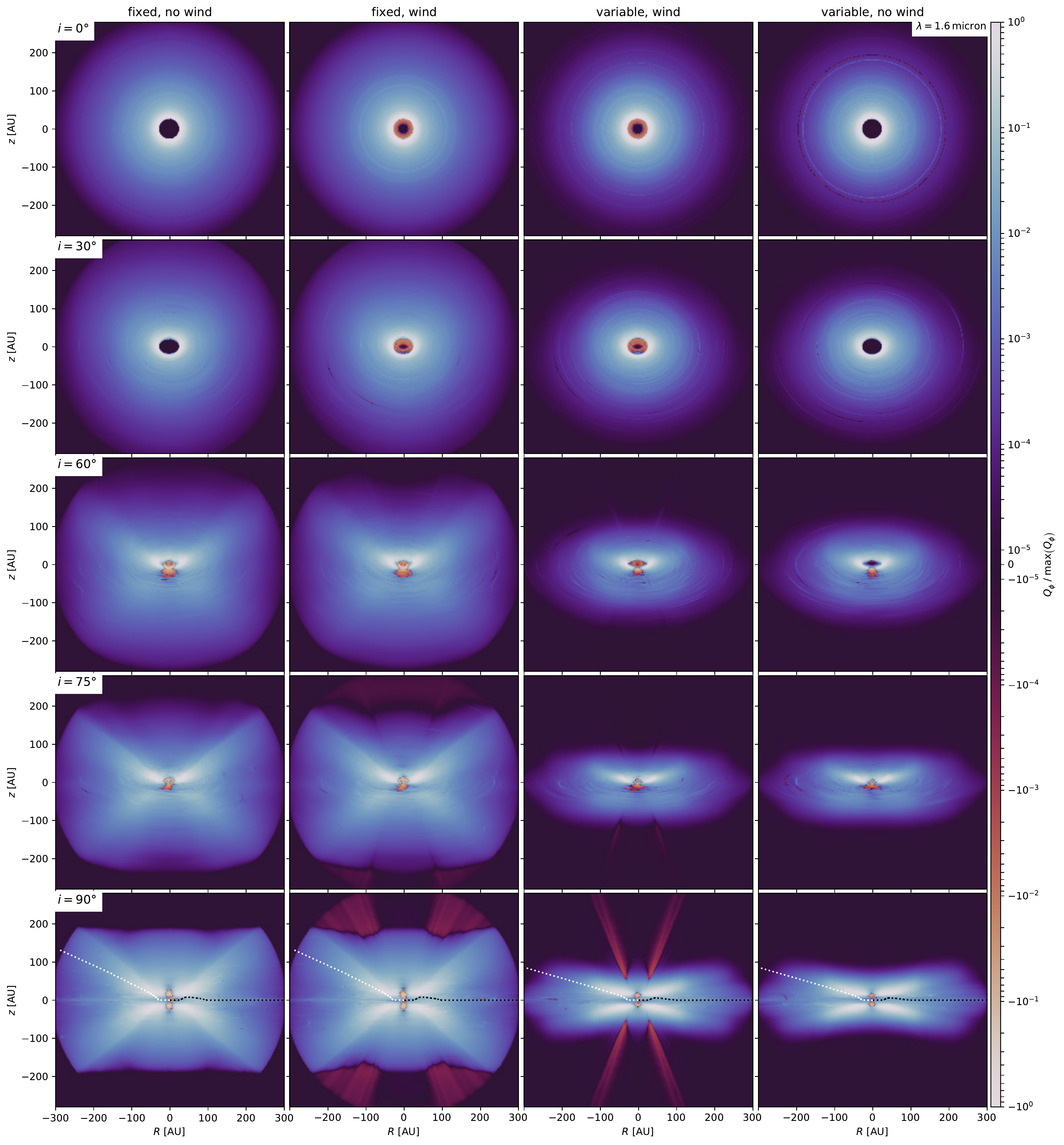}
    \caption{Polarised-light $Q_\phi$ for $\lambda_\mathrm{obs} = 1.6\,\mu$m for TD30, all else equal to Fig.~\ref{fig:radmc-qphi-20-1.2}.}
    \label{fig:radmc-qphi-30-1.6}
\end{figure*}

As noted in Sect.~\ref{sec:Discussion:pol}, reducing the instrument noise by a factor of 10 would result in a possible distinction between a `wind' and `no wind' disk with SPHERE IRDIS.
For reference, the corresponding synthesised image for TD20 (to be compared to Fig.~\ref{fig:irdis-20}) is shown in Fig.~\ref{fig:irdis-20-noise}.

\begin{figure*}
    \centering
    \includegraphics[width=0.99\textwidth]{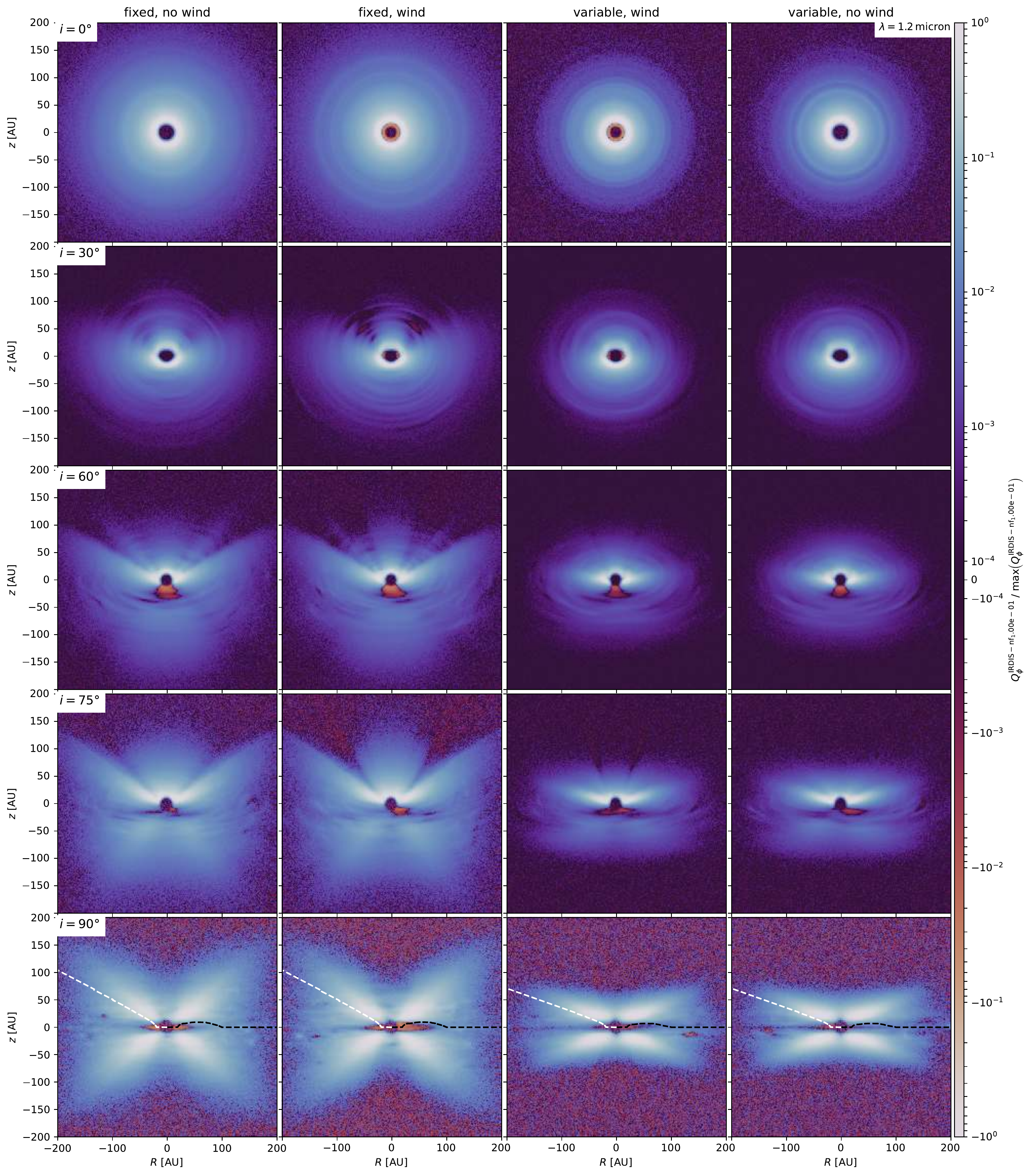}
    \caption{Synthesised instrument response for TD20 for SPHERE IRDIS in $J$-band, assuming a reduction of the instrument noise by a factor of 10; all else equal to Fig.~\ref{fig:irdis-20}.
    In contrast to said Fig.~\ref{fig:irdis-20}, there are visible differences between the `wind' and `no wind' models, for both fixTD20 and varTD20.}
    \label{fig:irdis-20-noise}
\end{figure*}

Furthermore, in Sect.~\ref{sec:Results:pol:irdis}, we investigated the difference $\Delta \alpha$ between the spectral indices $\alpha \equiv \alpha_{J,H}$ of the $P$-values of the `wind' and `no wind' models of TD20.
For reference, the individual spectral indices are included in Fig.~\ref{fig:irdis-20-alpha}; the noise masking of Fig.~\ref{fig:irdis-20-DeltaAlpha} has not been applied.

\begin{figure*}
    \centering
    \includegraphics[width=0.99\textwidth]{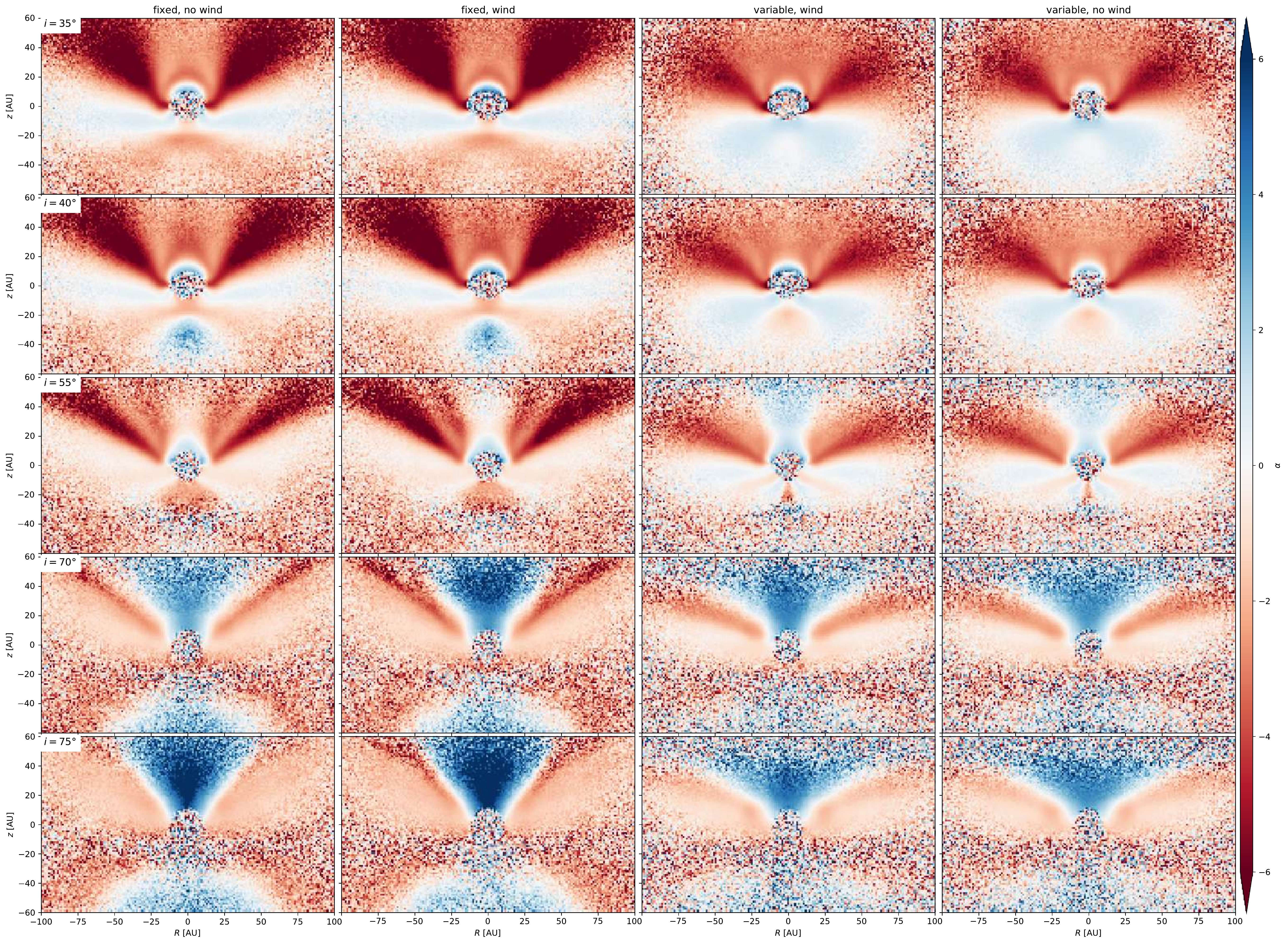}
    \caption{Spectral indices $\alpha \equiv \alpha_{J,H}$ for the individual models of TD20, at the inclinations shown in Fig.~\ref{fig:irdis-20-DeltaAlpha}.}
    \label{fig:irdis-20-alpha}
\end{figure*}

}

\end{appendix}

%%%%%%%%%%%%%%%%%%%%%%%%%%%%%%%%%%%%%%%%%%%%%%%%%%%%%%%%%%%%%%%%%%%%%%%%%%%%%%%%%%%%%%%%%%%%%%%%%%%%

\end{document}